\documentclass[aps,prb,twocolumn,amsmath,amssymb,nofootinbib,
  superscriptaddress]{revtex4-2}

\usepackage{graphicx}
\usepackage{bm}
\usepackage{amsmath}
\usepackage{amssymb}
\usepackage{physics}
\usepackage[colorlinks=true,linkcolor=blue,urlcolor=blue,citecolor=blue]{hyperref}
\usepackage[usenames,svgnames]{xcolor}
\usepackage{makecell}
\usepackage[normalem]{ulem}

\newcommand{\gcc}[1]{\textcolor{black}{#1}}

\newcommand{\Hop}{\hat{H}}

\newcommand{\Himp}{\hat{H}_{\rm imp}}
\newcommand{\gHimp}{\mathcal{H}_{\rm imp}}
\newcommand{\Hint}{\hat{H}_{\rm int}}

\newcommand{\aop}{\hat{a}}
\newcommand{\adop}{\hat{a}^{\dagger}}

\newcommand{\nop}{\hat{n}}

\newcommand{\cop}{\hat{c}}
\newcommand{\cdop}{\hat{c}^{\dagger}}

\newcommand{\Zbath}{Z_{\mathrm{bath}}}

\newcommand{\gK}{\mathcal{K}}
\newcommand{\gI}{\mathcal{I}}
\newcommand{\bolda}{\bm{a}}
\newcommand{\boldabar}{\bar{\bm{a}}}
\newcommand{\abar}{\bar{a}}

\newcommand{\gA}{\mathcal{A}}

\newcommand{\etabar}{\bar{\eta}}

\newcommand{\ustc}{State Key Laboratory of Precision and Intelligent Chemistry, University of Science and Technology of China, Hefei 230026, China}
\newcommand{\nudt}{Institute for Quantum Science and Technology, College of Science, National University of Defense Technology, Changsha 410073, China}

\begin{document}

\title{Scalable tensor network algorithm for quantum impurity problems}


\author{Zhijie Sun}
\affiliation{\ustc}

\author{Ruofan Chen}
\affiliation{College of Physics and Electronic Engineering, and Center for Computational Sciences, Sichuan Normal University, Chengdu 610068, China}

\author{Zhenyu Li}
\email{zyli@ustc.edu.cn}
\affiliation{\ustc}

\author{Chu Guo}
\email{guochu604b@gmail.com}

\affiliation{\nudt}



\begin{abstract}
The Grassmann time-evolving matrix product operator method has shown great potential as a general-purpose quantum impurity solver, as its numerical errors can be well-controlled and it is flexible to be applied on both the imaginary- and real-time axis. However, a major limitation of it is that its computational cost grows exponentially with the number of impurity flavors. In this work, we propose a multi-flavor extension of it to overcome this limitation. The key insight is that to calculate multi-time correlation functions on one or a few impurity flavors, one could integrate out the degrees of freedom of the rest flavors before hand, which could greatly simplify the calculation. The idea is particularly effective for quantum impurity problems with diagonal hybridization function, i.e., each impurity flavor is coupled to an independent bath, a setting which is commonly used in the field. We demonstrate the accuracy and scalability of our method for the imaginary time evolution of impurity problems with up to three impurity orbitals, i.e., $6$ flavors, and benchmark our results against continuous-time quantum Monte Carlo calculations. Our method paves the way of scaling up tensor network algorithms to solve large-scale quantum impurity problems. 
\end{abstract}

\maketitle


\section{Introduction}

The dynamical mean field theory (DMFT) has been a major effective numerical tool to solve strongly correlated quantum many-body problems~\cite{GeorgesRozenberg1996,GeorgesKotliar1992,MetznerVollhardt1989}. Instead of solving the original lattice problem, DMFT iteratively solves a quantum impurity problem (QIP) instead, in which an impurity is coupled to a noninteracting bath. In the past two decades, DMFT has been increasingly applied to solve realistic chemical or material problems beyond simple lattice models~\cite{KotliarMarianetti2006}. In these realistic situations, larger impurities have to be considered, and the scalability of the underlying quantum impurity solver basically determines the application range of DMFT.

The continuous-time quantum Monte Carlo (CTQMC) methods have been the golden standard for quantum impurity solvers~\cite{RubtsovLichtenstein2004,RubtsovLichtenstein2005,WernerMillis2006,GullTroyer2008,ChanMillis2009,haule2010-dynamical,GullWerner2011,huang2014-electronic,lu2016-pressure,yue2021-pairing}. They can yield numerically exact results, and are highly efficient on the imaginary-time axis, especially when the sign problem is not serious. However, real-time calculations remain a great challenge for CTMQC. A number of alternatives are also available, which complement CTQMC on the real-time axis or at zero temperature, including exact diagonalization~\cite{CaffarelKrauth1994,KochGunnarsson2008,GranathStrand2012,LuHaverkort2014,ZaeraLin2020,LuHaverkort2019,HeLu2014,HeLu2015}, numerical renormalization group~\cite{Wilson1975,Bulla1999,BullaPruschke2008,Frithjof2008,ZitkoPruschke2009,DengGeorges2013,StadlerWeichselbaum2015,LeeWeichselbaum2016,LeeWeichselbaum2017,CornagliaGrempel2004,PaaskeFlensberg2005,CornagliaNess2005,LaaksoMeden2014}, hierarchical equation of motion (HEOM)~\cite{YoshitakaKubo1989,jin2007-dynamics,jin2008-exact,yan2016-dissipation,cao2023-recent,ShiXu2018,YanShi2021,DanShi2023}, time-evolving matrix product state (MPS)~\cite{WolfSchollwock2014b,GanahlEvertz2014,GanahlVerstraete2015,WolfSchollwock2015,GarciaRozenberg2004,NishimotoJeckelmann2006,WeichselbaumDelft2009,BauernfeindEvertz2017,WernerArrigoni2023,KohnSantoro2021,KohnSantoro2022}. Unlike CTQMC, these methods generally suffer from various sources of \textit{uncontrolled numerical errors}, which could prohibit them from being used as general-purpose quantum impurity solvers.
\gcc{For example, the time-evolving MPS method unavoidably suffers from the bath discretization error as the bath is explicitly taken into account, it is also difficult to treat finite temperature baths~\cite{KohnSantoro2021}, or scale up to large impurities due to the fast entanglement growth in the impurity-bath wave function~\cite{WolfSchollwock2015}. HEOM, in comparison, is free of the bath discretization error, but it truncates the coupled operator equations to a certain order which could result in uncontrolled error, and its computational cost grows exponentially with the order. MPS has also been used in a development of HEOM to avoid the exponential cost growth~\cite{ShiXu2018}, but its potential to solve large impurity problems is yet to be demonstrated.}

The Grassmann time-evolving matrix product operator (GTEMPO) method, proposed by some of us, is a promising candidate as a general-purpose quantum impurity solver~\cite{XuChen2024}. GTEMPO represents the integrand of the impurity path integral (PI), referred to as the augmented density tensor (ADT), as a Grassmann MPS (GMPS), in analogous to the time-evolving matrix product operator (TEMPO) method for bosonic impurity problems~\cite{StrathearnLovett2018}. As the bath degrees of freedom are integrated out before hand via the Feynman-Vernon influence functional (IF)~\cite{FeynmanVernon1963}, GTEMPO is naturally free of the bath discretization error, similar to CTQMC. In fact, there are essentially only two sources of numerical errors in GTEMPO, one from discretization of the PI, the other from the MPS bond truncation. The first could be suppressed by using a finer time step size, while the second can be suppressed by using a larger bond dimension for MPS in principle. Meanwhile, as the core algorithms of GTEMPO only rely on the analytical expression of the IF, and that MPS algorithms are naturally free of the sign problem, GTEMPO can be flexibly applied on the imaginary-time axis~\cite{ChenGuo2024b}, real-time axis (i.e., Keldysh contour)~\cite{ChenGuo2024a,ChenGuo2024c}, or even the L-shaped Kadanoff contour~\cite{ChenGuo2024g,SunGuo2025}. \gcc{One could even extend GTEMPO to study impurity problems with dissipative baths (which has been studied using methods such as the stochastic Schr$\ddot{\text{o}}$dinger equation in the bosonic case~\cite{OrthLe2013,KamarMaghrebi2024}), as long as the bath remains quadratic and integrable, i.e., such that the hybridization function can be easily calculated.}

However, a major drawback of GTEMPO is that its computational cost grows exponentially with the impurity size~\cite{ChenGuo2024b}. This could be understood from its formalism: in GTEMPO, the IF is similar to the partition function of a 1D long-range classical Hamiltonian on the temporal axis, if the impurity is small enough (e.g., a single flavor) to be treated as 0D. However, for a large number of impurity flavors, the impurity is no longer 0D, and the Hamiltonian becomes quasi-2D with a non-negligible spatial dimension. From the general experience of treating quasi-2D problems with MPS algorithms, the computational cost (or more concretely the MPS bond dimension) would grow exponentially with the size of the shorter side~\cite{Schollwock2011}.
i.e, the number of flavors in our case.

In this work we propose a multi-flavor extension of the GTEMPO method to resolve this issue. 
As an impurity solver in DMFT, the central observables to calculate are the single-particle Green's functions which depends at most on two flavors. 
Therefore the degrees of freedom of the rest flavors can be integrated out before hand, similar to the treatment of the bath (although the flavor degrees of freedom can only be traced out numerically in general).
Even though the full ADT may not be compressed, it is highly possible that the reduced ADT, which keeps only the information of the involved flavors, can be largely compressed without significant loss of accuracy. This idea is implemented in this work on top of GTEMPO, referred to as the \textit{multi-flavor GTEMPO} method in the following. 
\gcc{We note that a similar idea has been implemented in the bosonic case, based on TEMPO~\cite{FuxKeeling2023}.}
We apply this method to study imaginary-time evolution of QIPs with up to three orbitals \gcc{(i.e., $6$ flavors, as a single electron orbital contains two flavors: spin up and down)}, and benchmark our results against CTQMC. Our numerical results show that this method is indeed effective: under a commonly used semi-circular spectrum function of the bath, we observe that we can obtain highly accurate results (the error is of the same order with the chosen time step size) with bond dimensions $200$ and $700$ in the two-orbital and three-orbital cases respectively, which are far smaller than those from exponential growth. 
These results illustrate the power of the multi-flavor GTEMPO method to solve large-scale QIPs. 

\gcc{To this end, we stress that although GTEMPO and the conventional time-evolving MPS methods both make use of MPS and are free of the sign problem, they are sharply different: the former represents the multi-time degrees of freedom of the impurity on the time axis, i.e., the ADT, as an MPS, while the latter represents the impurity-bath wave function at a certain time as an MPS. Therefore GTEMPO does not suffer from bath discretization error and is easy to treat finite temperature baths, GTEMPO is also likely to be more scalable to large impurity problems, especially with the development of the current work.}

The paper is organized as follows. In Sec.~\ref{sec:method}, we present the multi-flavor GTEMPO method. In Sec.~\ref{sec:results}, we show our numerical results ranging from single-orbital to three-orbital QIPs, together with converged DMFT iterations, and benchmark them with CTQMC calculations. We summarize in Sec.~\ref{sec:summary}.


\section{Method}\label{sec:method}

\subsection{Model Hamiltonian}
The Hamiltonian of the QIP can be generally written as
\begin{align}
\Hop = \Himp + \Hint,
\end{align}
where $\Himp$ denotes the impurity Hamiltonian, which can be generally written as
\begin{align}\label{eq:Himp}
\Himp = \sum_{p,q}t_{pq} \adop_p\aop_q + \sum_{p,q,r,s}g_{pqrs}\adop_p\adop_q\aop_r\aop_s.
\end{align}
Here $p,q,r,s$ are the fermion flavor labels that could contain both the spin and orbital indices, $\adop_p$ and $\aop_p$ are the fermionic creation and annihilation operators of the
$p$th impurity flavor, $t_{pq}$ is the tunneling strength and $g_{pqrs}$ is the interaction strength. 
$\Hint$ contains all the effects of the bath on the impurity.
In this work, we consider $\Hint$ in the following form:
\begin{align}\label{eq:Hint}
\Hint = \sum_{p, k} \epsilon_{p, k} \cdop_{p,k}\cop_{p,k} + \sum_{p, k}v_{p,k}(\adop_p\cop_{p,k} + \cdop_{p,k}\aop_p),
\end{align}
where $\cdop_{p, k}$ and $\cop_{p, k}$ are the fermionic creation and annihilation operators of the bath that is coupled to the $p$th impurity flavor, $\epsilon_{p, k}$ is the band energy and $v_{p,k}$ is the coupling strength between the impurity and bath. The first term on the right hand side of Eq.(\ref{eq:Hint}) is the free bath Hamiltonian and the second term is the coupling between the impurity and bath. 
We note that here we have restricted to diagonal coupling between impurity and bath, as each impurity flavor is coupled to its own bath in Eq.(\ref{eq:Hint}).
We will further assume $v_{p,k}=v_k$ for brevity (which does not lead to any significant simplification for GTEMPO).
The effect of $\Hint$ on the impurity dynamics is completely determined by the spectrum function, defined as $J(\epsilon) = \sum_k v_k^2 \delta (\epsilon -
  \epsilon_k)$.
Throughout this work, we will consider a semi-circular spectrum function (which is also used, for example, in Refs.~\cite{WolfSchollwock2014,WolfSchollwock2015})
\begin{align}\label{eq:Jw}
    J(\epsilon) = \frac{2}{\pi D} \sqrt{1-(\epsilon/D)^2},
\end{align}
in which we set $D=2$ and use it as the unit. We will focus on the imaginary-time evolution in this work, but we stress that our method can be straightforwardly applied to the Keldysh or Kadanoff contours as well.

\subsection{The multi-flavor GTEMPO method}

The starting point of the GTEMPO method is the impurity path integral, which can be written as
\begin{align}\label{eq:Z}
  Z = \Zbath \int \mathcal{D} [\boldabar\bolda] \gK [\boldabar \bolda]\gI [\boldabar \bolda]
\end{align}
where $\Zbath$ is the partition function of the bath, $\boldabar_p = \{\abar_p(\tau)\}$, $\bolda_p = \{a_p(\tau)\}$ are the Grassmann trajectories for the $p$th flavor over the whole imaginary-time interval $[0,\beta]$ ($\beta$ is the inverse temperature), and $\boldabar = \{\boldabar_p, \boldabar_q, \cdots\}$, $\bolda = \{\bolda_p, \bolda_q, \cdots\}$. 
The measure $\mathcal{D} [\boldabar\bolda]$ is 
\begin{align}
\mathcal{D} [\boldabar\bolda] = \prod_{p, \tau} \dd\abar_p(\tau)\dd a_p(\tau) e^{-\abar_p(\tau)a_p(\tau)}.
\end{align}
$\gK [\boldabar \bolda]$ denotes the contribution from the bare impurity dynamics determined by $\Himp$ only, which can be formally written as
\begin{align}\label{eq:K}
\gK[\boldabar \bolda] = e^{-\int_{0}^{\beta} \dd \tau\gHimp(\tau)},
\end{align}
where $\gHimp(\tau)$ is obtained from $\Himp$ by making the substitutions $\aop_p\rightarrow a_p(\tau)$, $\adop_p\rightarrow \abar_p(\tau)$.
$\gI [\boldabar \bolda]$ is the Feynman-Vernon IF that is determined by $\Hint$ only. For diagonal hybridization function, the IF can be calculated as $\gI [\boldabar \bolda] = \prod_p \gI_p [\boldabar_p \bolda_p]$ with $\gI_p [\boldabar_p \bolda_p]$ the IF of the $p$th flavor, which can be explicitly written as
\begin{align}\label{eq:IF}
\gI_p [\boldabar_p \bolda_p] = e^{-\int_0^{\beta}\dd\tau'\int_0^{\beta}\dd\tau''\abar_p(\tau')\Delta(\tau', \tau'')a_p(\tau'')}.
\end{align}
The function $\Delta(\tau', \tau'')$ in Eq.(\ref{eq:IF}) is usually referred to as the hybridization function, which directly determines the IF, and can be calculated \gcc{as~\cite{ChenGuo2024b}:
\begin{align}
\Delta(\tau', \tau'') = \int \dd\epsilon J(\epsilon)D_{\epsilon}(\tau', \tau''),
\end{align}
where $D_{\epsilon}(\tau', \tau'')$ is the free bath Matsubara Green's function defined as~\cite{Chen2025}
\begin{align}
D_{\epsilon}(\tau', \tau'') = -[\Theta(\tau' - \tau'') - n(\epsilon)]e^{-\epsilon(\tau' - \tau'')},
\end{align}
with $\Theta$ the Heaviside step function and $n(\epsilon) = (e^{\beta\epsilon}+1)^{-1}$ the Fermi–Dirac distribution.
}
The integrand in Eq.(\ref{eq:Z}) defines the \textit{augmented density tensor}, denoted as
\begin{align}\label{eq:adt}
\gA [\boldabar\bolda]= \gK [\boldabar\bolda] \prod_p\gI_p [\boldabar_p\bolda_p],
\end{align}
which contains all the information of the multi-time impurity dynamics. 
Here we point out that the fermionic ADT is essentially the same as a recently proposed concept: the process tensor~\cite{JorgensenPollock2019,GuoPoletti2020,ZhangGuo2025,TarantoModi2025}, except that the ADT is often discussed in the context of impurity problems.

\begin{figure}
  \includegraphics[width=\columnwidth]{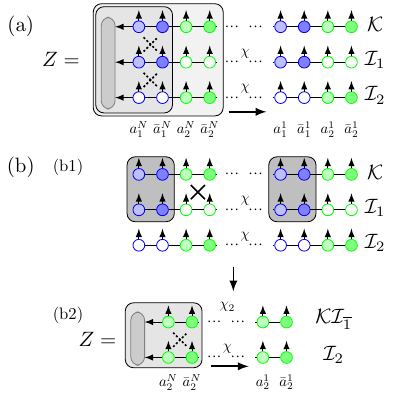} 
  \caption{(a) Schematical illustration of the zipup algorithm to to calculate the partition function for an impurity problem with $2$ flavors, where the quasi-2D tensor network is contracted (integrating out the pairs of conjugate Grassmann variables) from left to right and the augmented density tensor $\gA$ is calculated on the fly as indicated by the dashed $\times$. (b) The scheme used in the multi-flavor GTEMPO method to calculate the partition function based on the reduced ADT $\gA_2$ for the second flavor, in which one first calculate $\gK\gI_{\overline{1}}$ by multiplying $\gK$ and $\gI_1$ and integrating out the first flavor (b1), and then multiply $\gK\gI_{\overline{1}}$ and $\gI_2$ to obtain $\gA_2$ (b2). Again the second multiplication is only performed on the fly similar to (a). The empty circles in (b1) mean that these Grassmann variables do not exist in the corresponding GMPS, while the gray box means that the Grassmann variables inside it will be integrated out after multiplication.
    }
    \label{fig:demo}
\end{figure}

In GTEMPO, the continuous Grassmann trajectories $\boldabar_p$, $\bolda_p$ are first discretized into $N$ discrete Grassmann variables (GVs) with equal-distant imaginary-time step size $\delta\tau=\beta/N$, then $\gK$ and $\gI_p$ are systematically constructed as GMPSs (one could see Ref.~\cite{ChenGuo2024b} for those details). 
The bond dimension of $\gK$ is generally a constant determined by the number of flavors, denoted as $n$, while the bond dimension of each $\gI_p$, denoted as $\chi$ in the following, will usually be larger than that of $\gK$.
With $\gK$ and $\gI_p$, one could in principle multiply them together to obtain $\gA$ as a GMPS as in the definition of Eq.(\ref{eq:adt}) (the multiplication of two GMPSs originates from the multiplication of two Grassmann tensors, which is analogous to the element-wise multiplication between two normal tensors~\cite{GuoChen2024d}). 
Based on $\gA$, one can easily compute any multi-time impurity correlation functions. For example, the Matsubara Green's function can be calculated as
\begin{align}\label{eq:Matsubara}
G_{pq}(j\delta\tau) = \langle \aop_p(j\delta\tau)\adop_q(0)\rangle = Z^{-1}\int\mathcal{D} [\boldabar\bolda] a_{p}^j\abar_{q}^0 \gA [\boldabar \bolda],
\end{align}
where $\langle \hat{X}\rangle $ means the thermal average of operator $\hat{X}$.
However, the bond dimension of $\gA$ will grow exponentially with $n$ in general, or more concretely, as $O(\chi^{n})$ (and the computational cost will grow as $O(\chi^{2n})$). 
A more efficient approach is to use the zipup algorithm, first proposed in Ref.~\cite{ChenGuo2024a}, which directly contracts the quasi-2D tensor network made of $\gK$ and $\gI_p$ using a left to right sweep, during which $\gA $ is calculated on the fly (this strategy is similar to exactly contracting a 2D tensor network from the shorter side~\cite{GuoWu2019}). The zipup algorithm could reduce the computational cost down to $O(\chi^n)$, which is illustrated in Fig.~\ref{fig:demo}(a). However, the cost is still exponential.

In the multi-flavor GTEMPO method, we resolve this exponential growth by realizing that to calculate $G_{pq}(\tau)$, one only needs the information of flavors $p$, $q$, and all the rest flavors can be integrated out before hand. We denote $\gA_{pq}[\boldabar_{pq} \bolda_{pq} ]$ as the \textit{reduced augmented density tensor} that only contains the $p$, $q$ flavors, which can be calculated from $\gA[\boldabar \bolda]$ as:
\begin{align}\label{eq:reducedadt}
\gA_{pq}[\boldabar_{pq} \bolda_{pq}] = \int \mathcal{D}[\boldabar_{\overline{pq}} \bolda_{\overline{pq}}] \gA [\boldabar \bolda].
\end{align}
Here $\overline{pq}$ means that all the flavors are included except $p,q$. With $\gA_{pq}$, we can calculate $G_{pq}(\tau)$ as
\begin{align}\label{eq:Matsubara2}
G_{pq}(\tau)  = Z^{-1}\int\mathcal{D} [\boldabar_{pq} \bolda_{pq}] a_{p}^j\abar_{q}^0 \gA_{pq} [\boldabar_{pq} \bolda_{pq}].
\end{align}

In the following, we will show our detailed numerical scheme to calculate the reduced ADT $\gA_{pq}$ (in fact, we will only calculate it on the fly as in the original zipup algorithm). Here we will focus on the case $p=q=n$ and denote $\gA_{pq}=\gA_{n}$, $G_{pq}=G_n$, but the procedure can be straightforwardly generalized to arbitrary $p$ and $q$.


First, we multiply $\gK$ with $\gI_1$, then we can integrate out the first flavor as the rest $\gI_p$s contain no information of it, the result is denoted as $\gK\gI_{\overline{1}}[\boldabar_{\overline{1}}\bolda_{\overline{1}}]$:
\begin{align}\label{eq:KI1}
\gK\gI_{\overline{1}}[\boldabar_{\overline{1}}\bolda_{\overline{1}}] = \int \mathcal{D}[\boldabar_1\bolda_1] \gK [\boldabar\bolda] \gI_1 [\boldabar_1\bolda_1].
\end{align}
After that, if there is still more than one flavor left, we multiply $\gK\gI_{\overline{1}}$ with $\gI_2$ and integrate out the second flavor, and denote the result as $\gK\gI_{\overline{12}}$. Repeating this process, until that we have only $\gI_n$ left. In this stage, we can simply obtain $\gA_n$ by multiplying $\gK\gI_{\overline{1\cdots n-1}}$ and $\gI_n$, however, we opt to keep them both and directly use them to calculate observables on the $n$th flavor using zipup algorithm. This scheme is schematically illustrated in Fig.~\ref{fig:demo}(b) with the $n=2$ case. During this process, one needs to compress the intermediate GMPS (otherwise the bond dimension will still grow exponentially), i.e., the $\gK\gI_{\overline{1\cdots j}}$s ($1\leq j<n$), for which we use another bond dimension $\chi_2$. For large impurities, $\chi_2$ will generally be larger than $\chi$ and determines the computational cost of the multi-flavor GTEMPO method.

To this end, we note that our scheme will be effective only if $\chi_2 \ll O(\chi^n)$. We argue that this behavior of $\chi_2$ is expected. In fact, this scheme is similar to the boundary MPS method used to contract 2D tensor networks~\cite{VerstraeteCirac2004}, but with a crucial difference: a partial integration is performed immediately after each GMPS multiplication, during which we \textit{throw away} the information of the flavor being integrated out (the length of the resulting GMPS becomes shorter correspondingly). Therefore it is reasonable that the intermediate GMPSs can be significantly compressed without loss of accuracy. In comparison, it has been shown that if one directly contract the quasi-2D tensor network in Fig.~\ref{fig:demo}(a) in full analogous to the boundary MPS algorithm without doing partial integration, them compression of the intermediate GMPSs will result in significant numerical errors~\cite{ChenGuo2024b}. 

In the next section, we will illustrate the effect of $\chi_2$ on the numerical accuracy of $G(\tau)$ with various examples ranging from one to three orbitals.

\section{Numerical results}\label{sec:results}

In our numerical examples, we will focus on calculating $G_{pq}(\tau)$ with $p=q=n$ and neglect the subscript. The CTQMC calculations in this work are performed using the TRIQS package~\cite{ParcolletSeth2015,SethParcollet2016} with $6.4 \times 10^9$ samples.
As we consider $\Hint$ that is symmetric for all the flavors, $\gI_p$ is the same for all flavors (which is not a significant simplification though, since the cost of calculating the reduced ADT dominates for large impurities), so can reuse the same $\gI_p$ across all our multi-flavor GTEMPO calculations for different number of flavors, as long as the bath parameters $\beta$ and $\delta \tau$ remain the same. 

In GTEMPO, there are only two sources of numerical errors: (1) the time discretization error of the impurity PI, characterized by $\delta\tau$, and (2) the MPS bond truncation error, occurred in building each $\gI_p$ as a GMPS, characterized by $\chi$. In multi-flavor GTEMPO, there is an additional source of error, occurred in compressing the intermediate GMPSs $\gK\gI_{\overline{1\cdots j}}$, characterized by $\chi_2$ as indicated in Fig.~\ref{fig:demo}(b).
In the following, we will first analyze the errors of GTEMPO, against $\chi$ and $\delta\tau$ in the single-flavor case in Sec.~\ref{sec:Toulouse}. 
Then we consider the single-orbital case in Sec.~\ref{sec:AIM} and the two- and three-orbital cases in Sec.~\ref{sec:moreorbitals}, where we analyze the errors of our multi-flavor GTEMPO against $\chi_2$. Finally in Sec.~\ref{sec:dmft}, we demonstrate the effectiveness of our multi-flavor GTEMPO with two- and three-orbital converged DMFT iterations.
We will use the mean error, defined as $\sqrt{\sum_{i=1}^N{{|x_i-y_i|}^2/N}}$
between two sets of results $\bold{x}$ and $\bold{y}$, to characterize the error between GTEMPO or multi-flavor GTEMPO results and other calculations. 

\subsection{Toulouse model}\label{sec:Toulouse}

\begin{figure}[!h]\centering
  \includegraphics[width=\columnwidth]{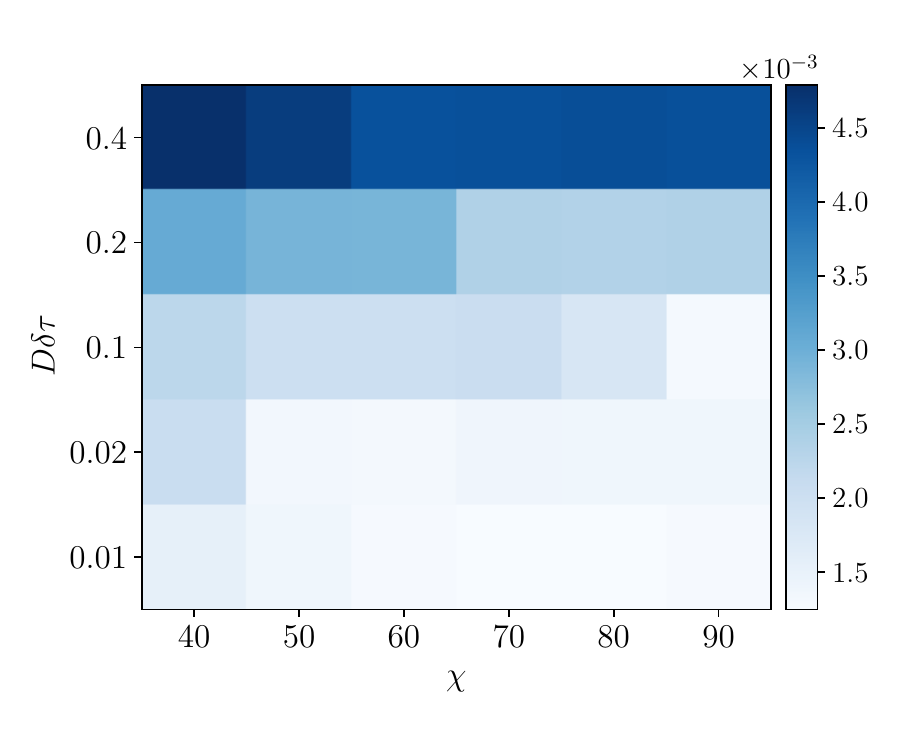}
  \caption{Mean error of $G(\tau)$ between GTEMPO and the analytical solution, as a function of $\delta \tau$ and $\chi$, for the Toulouse model with $\beta=10$ and $\epsilon_d = 1$.}
    \label{fig:toulouse}
\end{figure}

We first consider the single-flavor case with $\Himp = \epsilon_d\adop\aop$, referred to as the Toulouse model, which can be analytically solved~\cite{mahan2000-many}. We will fix $\epsilon_d=1$. For this model there is only a single $\gI_1$ and there is no need to use the multi-flavor GTEMPO method. The purpose of this \gcc{section} is merely to find the proper values of the two hyperparameters $\chi$ and $\delta \tau$, \gcc{which will be used in the multi-flavor GTEMPO calculations in later sections.}

In Fig.~\ref{fig:toulouse}, we show the mean error of $G(\tau)$ between GTEMPO and the analytical solution as a function of $\delta \tau$ and $\chi$, for the Toulouse model at $\beta = 10$. 
We observe that the accuracy systematically improves with smaller $\delta \tau$ and larger $\chi$. In particular, with $\delta \tau=0.05$ and $\chi=60$ GTEMPO has already achieved an accuracy of $2.0\times 10^{-3}$. Interestingly, even with a large $\delta \tau=0.2$ and a small $\chi=40$, GTEMPO still achieves an accuracy of $4.8\times 10^{-3}$, which illustrates the robustness and accuracy of it.
In all the subsequent simulations, we will fix $\delta \tau = 0.05$.

\begin{figure}[!h]\centering
  \includegraphics[width=\columnwidth]{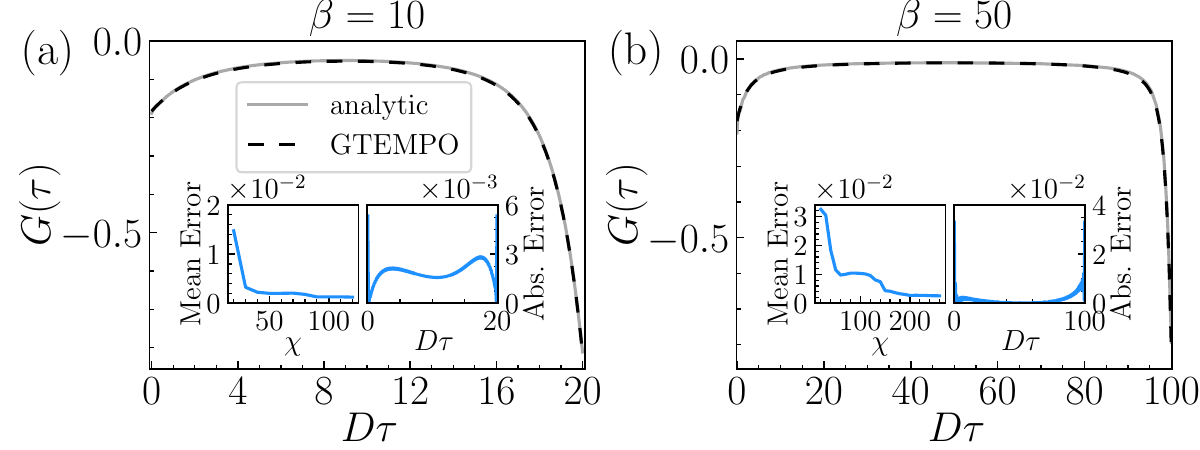}
  \caption{$G(\tau)$ for the Toulouse model at (a) $\beta = 10$ and (b) $\beta = 50$. The gray solid lines and the black dashed lines represent the analytical solutions and the GTEMPO results respectively.
  The left inset in both panels shows the mean error between GTEMPO and the analytical solution as a function of $\chi$, while the right inset shows the absolute error of $G(\tau)$ between GTEMPO and the analytical solution. We have used $\chi=60$ in the main panel (a) and its right inset, and used $\chi=200$ in the main panel (b) and its right inset.
  }
    \label{fig:toulouse2}
\end{figure}

In Fig.~\ref{fig:toulouse2}(a,b), we study the accuracy of GTEMPO results against different $\chi $ for the Toulouse model at $\beta = 10$ and $\beta = 50$, respectively. 
We can see that the GTEMPO results generally agree very well with the analytical solutions.
From the left insets, we can see that the mean error goes down monotonically with $\chi$ in both cases. From the insets of both panels, we can see that the errors are generally larger at $\beta=50$ compared to $\beta=10$, even though larger $\chi$s have been used, which indicates that GTEMPO has a larger computational cost at lower temperature.

\subsection{Single-orbital Anderson impurity model}\label{sec:AIM}



\begin{figure}[!h]\centering
  \includegraphics[width=\columnwidth]{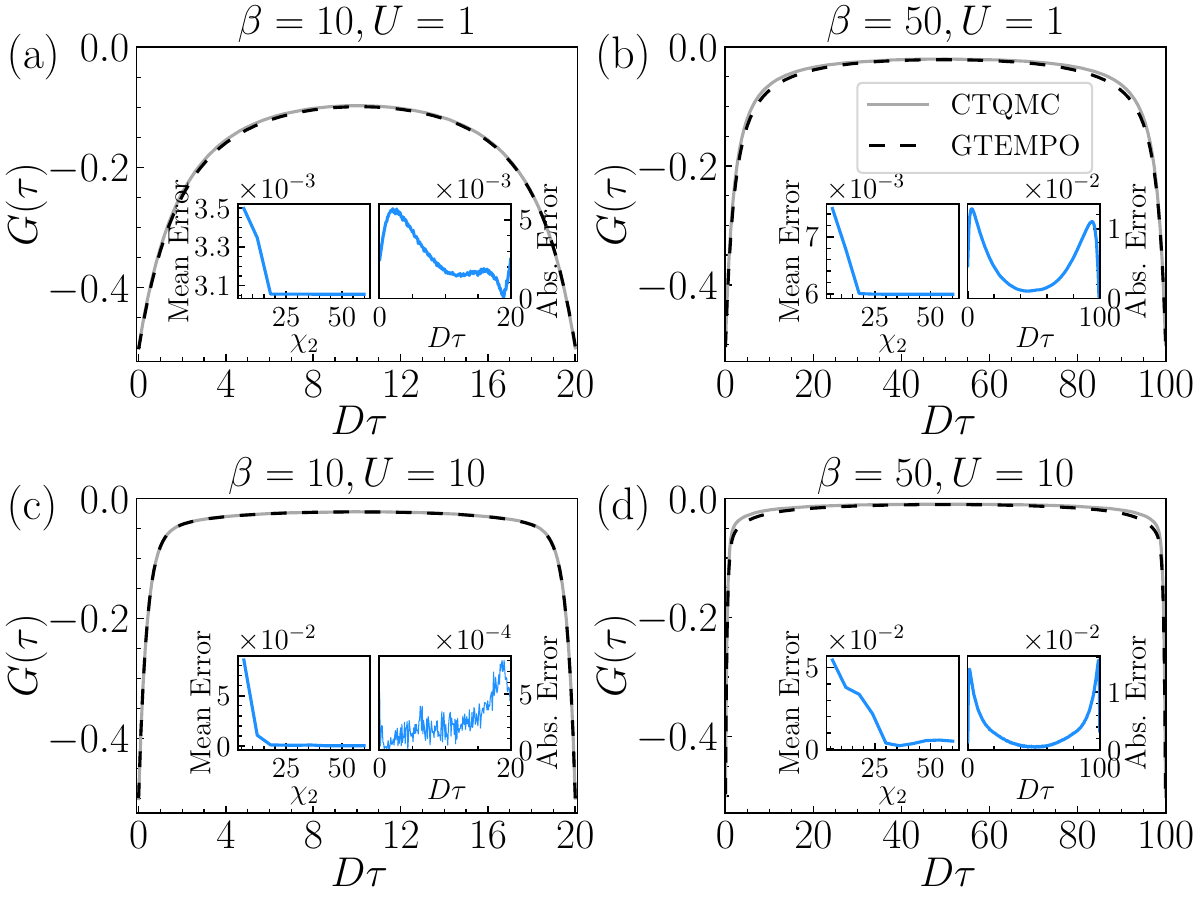}
  \caption{$G(\tau)$ as a function of $\tau$ for the single-orbital Anderson impurity model with impurity Hamiltonian in Eq.(\ref{eq:AIM}), under different parameter settings as shown in the titles. The black dashed lines and the gray solid lines represent the multi-flavor GTEMPO results and the CTQMC results respectively. The left inset in all panels shows the mean error between multi-flavor GTEMPO and CTQMC as a function of $\chi_2$, the right inset shows the absolute error as a function of $\tau$.
  We have used $\chi=60$ for all the simulations in (a,c), and $\chi=200$ for all the simulations in (b,d).
  For all the main panels, and their right insets, we have used $\chi_2=60$.
  }
    \label{fig:SIAM}
\end{figure}

Now we move on to the single-orbital case ($2$ flavors) where the multi-flavor GTEMPO method can be seriously tested. We consider the single-orbital Anderson impurity problem with 
\begin{align}\label{eq:AIM}
\Himp = \epsilon_d \sum_{p=1}^2 \nop_p + U\nop_1\nop_2,
\end{align}
where $\nop_p = \adop_p\aop_p$ is the electron density operator. We focus on the half-filling scenario with $\epsilon_d=-U/2$, which allows an easy justification of the results as $G(0)=G(\beta)=-0.5$ is exactly satisfied in this scenario.

In Fig.~\ref{fig:SIAM}, We consider four different parameter settings with different $\beta$ and $U$, as shown in the titles of the four panels, to illustrate the accuracy of the multi-flavor GTEMPO in different cases. We have used $\chi=60$ for $\beta=10$ and $\chi=200$ for $\beta=50$, and used $\chi_2=60$ for all the simulations in the main panels. 
We can see from the main panels that the multi-flavor GTEMPO results generally agree very well with CTQMC, and the error is the largest at $\beta=50$ and $U=10$ which is expected. From the left insets of all the panels, we observe that the errors between multi-flavor GTEMPO and CTQMC decrease very quickly with $\chi_2$ and well saturates with $\chi_2 = 30$. These results already illustrate the effectiveness of our method, since if we directly multiply $\gK$ and $\gI_1$, the bond dimension of the resulting GMPS will be $4\chi$ (the bond dimension of $\gK$ grows as $2^n$, which is $4$ for this model), much larger than $\chi_2$.
The right insets show the absolute errors between multi-flavor GTEMPO and CTQMC, which are on the order of $10^{-3}$ for $\beta=10$, and $10^{-2}$ for $\beta=50$.


\subsection{Multi-orbital quantum impurity models}\label{sec:moreorbitals}

\begin{figure}[!h]\centering
  \includegraphics[width=\columnwidth]{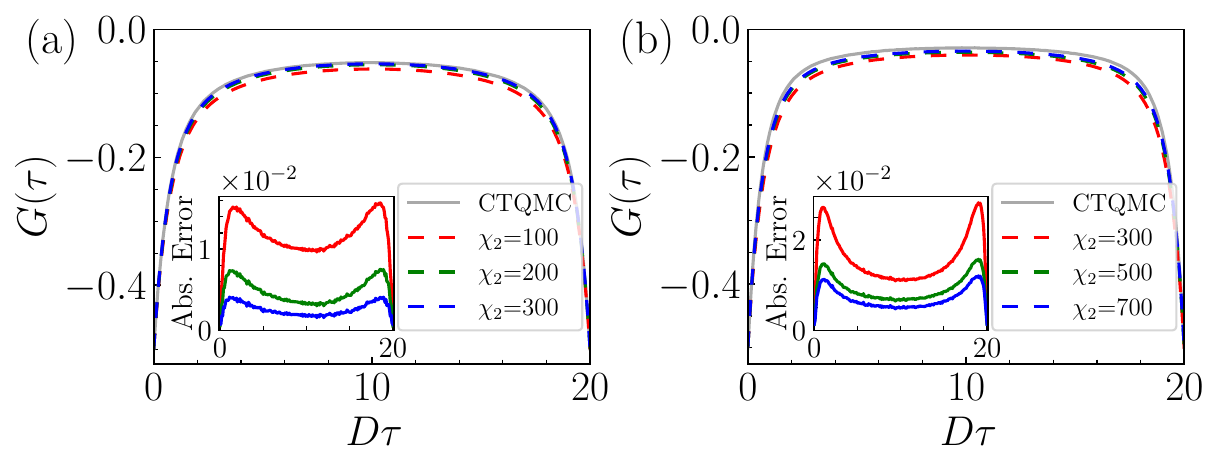}
  \caption{$G(\tau)$ for the two-orbital (a) and three-orbital (b) impurity problems with impurity Hamiltonian in Eq.(\ref{eq:Kanamori}). The gray solid lines in both panels represent the CTQMC results. The red, green and blue dashed lines represent multi-flavor GTEMPO results with $\chi_2=100,200,300$ in (a), and represent multi-flavor GTEMPO results with $\chi_2=300,500,700$ in (b).
  The insets show the absolute errors between multi-flavor GTEMPO and CTQMC results, where the colored solid line corresponds to multi-flavor GTEMPO result plotted in dashed line with the same color in the main panel.
  } 
    \label{fig:orb23_conv}
\end{figure}

In the next, we apply the multi-flavor GTEMPO method to study multi-orbital impurity problems, in which we will further fix $\beta=10$ and $\chi=60$ such that we can focus on analyzing the errors against $\chi_2$.
We consider multi-orbital impurity Hamiltonian of the Slater-Kanamori type, given by~\cite{GullWerner2011}
\begin{align}\label{eq:Kanamori}
    \hat{H}_{\mathrm{imp}} 
    &= \epsilon_d \sum_p \hat{a}_p^{\dagger} \hat{a}_p 
    + U \sum_x \hat{a}_{x, \uparrow}^{\dagger} \hat{a}_{x, \downarrow}^{\dagger} \hat{a}_{x, \downarrow} \hat{a}_{x, \uparrow} \nonumber \\
    &+ (U-2 J) \sum_{x \neq y} \hat{a}_{x, \uparrow}^{\dagger} \hat{a}_{y, \downarrow}^{\dagger} \hat{a}_{y, \downarrow} \hat{a}_{x, \uparrow} \nonumber \\ 
    &+ (U-3 J) \sum_{x>y, \sigma} \hat{a}_{x, \sigma}^{\dagger} \hat{a}_{y, \sigma}^{\dagger} \hat{a}_{y, \sigma} \hat{a}_{x, \sigma} \nonumber \\
    &- J \sum_{x \neq y}\left(\hat{a}_{x, \uparrow}^{\dagger} \hat{a}_{x, \downarrow}^{\dagger} \hat{a}_{y, \uparrow} \hat{a}_{y, \downarrow}+\hat{a}_{x, \uparrow}^{\dagger} \hat{a}_{y, \downarrow}^{\dagger} \hat{a}_{y, \uparrow} \hat{a}_{x, \downarrow}\right).
\end{align}
Again we will focus on the half-filling scenario, with conditions
\begin{align}
  \epsilon_d &= -(3U - 5J)/2; \\
  \epsilon_d &= -(2.5U - 5J),
\end{align}
for these two cases respectively~\cite{Sherman2020,WernerMillis2009}.
In our numerical simulations, we fix $U=4$ and $J=1$, and $\epsilon_d$ is set by the half-filling conditions. 
\gcc{Here we note that with the spectrum function in Eq.(\ref{eq:Jw}), we have chosen a bandwidth $4$ and a coupling strength between the impurity and bath to be $1$. As the energy scale of usual materials is eV, if we take the coupling strength as $1 $ eV, which is about $11605$ K, then $\beta = 10$ corresponds to $\sim1160$ K, which is about $4$ times higher than room temperature.}

In Fig~\ref{fig:orb23_conv}(a,b), we plot $G(\tau)$ as a function of $\tau$ for the two- and three-orbital cases respectively, where the gray solid lines are CTQMC results, the red, green and blue dashed lines are multi-flavor GTEMPO results calculated with different $\chi_2$s. In the insets, we plot the absolute errors between multi-flavor GTEMPO and CTQMC. We can see that the absolute errors are of the order $10^{-2}$ in both cases, and decrease monotonically with larger $\chi_2$. 
These results clearly illustrate the power of the multi-flavor GTEMPO method: we can already obtain very accurate results with absolute error around or smaller than $10^{-2}$, even though the $\chi_2$ used in these two cases (which are $300$ and $700$ at most) are far smaller than those required by exponential growth, which are $16\times 60^3\approx 3.5\times10^6$ and $64\times 60^5\approx 5\times 10^{10}$ respectively! 

\gcc{In fact, in Ref.~\cite{ChenGuo2024b}, similar two- and three-orbital impurity models have been studied using vanilla GTEMPO, but only $\beta=10$ and $\beta=2$ have been reached for these two cases respectively due to the exponential growth of cost. With multi-flavor GTEMPO, we still expect an exponential growth of $\chi_2$ with the number of flavors for general impurity problems, but of the form $O(2^n\chi)$ instead of $O(2^n\chi^n)$. The reason is that in multi-flavor GTEMPO, we believe the influence of the rest flavors on the current flavor, after been integrated out, is similar to adding in some effective interaction among the flavors, and if this is the case, the largest contribution it can have on the bond dimension of the ADT is $O(2^n)$ (we note that in the bosonic version of this method, a large impurity with $21$ spins has been studied~\cite{FuxKeeling2023}).
Under this scaling, we would expect $\chi_2\approx 3000$ and $\chi_2\approx 10000$ are sufficient for the four- and five-orbital cases respectively, under the same $\beta$.
In addition, it has been observed in Ref.~\cite{ChenGuo2024b} that the bond dimension $\chi$ of $\gI$ grows approximately linearly with the inverse temperature. Therefore we would also expect a linear growth of $\chi_2$ for multi-flavor GTEMPO against $\beta$.}


\subsection{DMFT iterations on the Bethe lattice}\label{sec:dmft}

\begin{figure}[!h]\centering
  \includegraphics[width=\columnwidth]{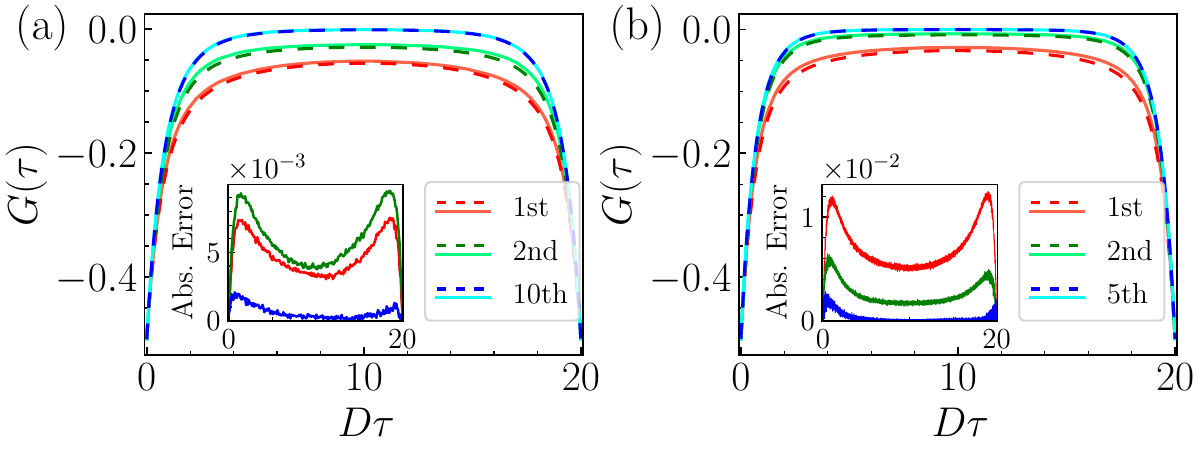}
  \caption{$G(\tau)$ at the 1st, 2nd, and final iterations in the DMFT calculation for the two-orbital (a) and three-orbital (b) DMFT iterations for the Slater-Kanamori model on the Bethe lattice. The dashed lines are the multi-flavor GTEMPO results at different iterations and the solid lines close to them are the corresponding CTQMC results. We have used $\chi_2=200$ in (a) and $\chi_2=700$ in (b). The insets show the absolute errors between multi-flavor GTEMPO and CTQMC results at different DMFT iterations.
  }
    \label{fig:orb23_DMFT}
\end{figure}

Finally, we apply our multi-flavor GTEMPO as an impurity solver in DMFT, to solve the multi-orbital Slater-Kanamori model~\cite{GullWerner2011} on the Bethe lattice. The hopping parameter of the Bethe lattice is chosen to be $1$.
To benchmark our results against CTQMC in a step-by-step fashion, we simply choose Eq.(\ref{eq:Jw}) as the initial spectral function for both solvers.
We have performed $10$ DMFT iterations in the two-orbital case with $\chi_2=200$, and $5$ DMFT iterations in the three-orbital case with $\chi_2=700$. 
In both case the DMFT iterations have well converged, with mean errors between the last two iterations to be $5.5\times 10^{-5}$ and $3.7\times 10^{-4}$. 
The results of the two cases are plotted in Fig.~\ref{fig:orb23_DMFT}(a,b) respectively. In the insets we show the absolute errors between multi-flavor GTEMPO and CTQMC results. 
We can see that the absolute errors in both cases, and at different iterations, are all of the order $10^{-2}$ or less, and interestingly, the errors in the last DMFT iteration is the smallest in both cases.
These results thus illustrate the power of the multi-flavor GTEMPO method as a scalable quantum impurity solver.


\gcc{To this end, we briefly discuss the errors in multi-flavor GTEMPO as well as in general DMFT applications. 
The first source of error in multi-flavor GTEMPO is the time discretization error, determined by the discrete time step size, e.g., $\delta\tau$ for imaginary-time calculations. In our numerical simulations we have mostly used $\delta\tau=0.05$ and observed an absolute error of the order $10^{-2}$. 
The time discretization error can be systematically suppressed by using a smaller time step size, or using higher-order Trotter decompositions of the PI~\cite{makarov1994-path,makri1995-numerical} or even continuous MPS~\cite{ParkChan2024}.
The second source of error in multi-flavor GTEMPO is the MPS bond truncation error, determined by $\chi$ and $\chi_2$. The MPS bond truncation error can be systematically suppressed by using larger $\chi$ and $\chi_2$. As a common practice in MPS algorithms~\cite{XiangLi2024}, one could even perform an extrapolation against $\chi$ and $\chi_2$ to the infinite bond dimension limit to eliminate this error. On the other hand, the DMFT procedure is often tolerant of small errors. For example, in early times the Hirsch-Fye algorithm was widely adopted as an impurity solver for DMFT applications~\cite{georges1996-dynamical}, which, besides the Monte Carlo sampling error, has a time discretization error of the order $\delta\tau^2$. The time step size $\delta\tau$ in this method is usually of the order $0.1$, and thus the discretization error is of order $10^{-2}$, similar to that in our simulations. In addition, CTQMC is widely adopted as a modern impurity solver~\cite{GullWerner2011}. When applied to nonequilibrium DMFT, its error can also reach the order of $10^{-2}$, for example, see Fig. 25 in Ref.~\cite{AokiWerner2014}.
Overall, the errors in GTEMPO can be well controlled and systematically suppressed, compared to its alternatives, but as an impurity solver in DMFT, one may need to balance the numerical accuracy and computational costs, especially for large-scale impurity problems.
}

\section{Summary}\label{sec:summary}
In summary, we have proposed a multi-flavor Grassmann time-evolving matrix product operator method, which builds on top of the GTEMPO method and particularly targets at a large number of impurity flavors. The key insight is that to calculate observables on a single or a few flavors, one could integrate out the degrees of freedom of other flavors, resulting in a much smaller reduced augmented density tensor than the full ADT. In our numerical examples ranging from single-orbital to three-orbital quantum impurity problems, we have shown that in the multi-flavor GTEMPO method, the growth of the bond dimension of the involved GMPSs is far slower than exponential which is required in vanilla GTEMPO. Under a commonly used semi-circular spectrum function of the bath, we observe that with bond dimensions $200$ and $700$ for the two-orbital and three-orbital cases, we can already obtain accurate Matsubara Green's functions whose absolute errors compared to CTQMC results are within the order of $10^{-2}$, we can also achieve converged DMFT iterations in which the multi-flavor GTEMPO results match the CTQMC results step by step.

In addition, we note that the multi-flavor GTEMPO method can be straightforwardly applied on the real-time axis, similar to the GTEMPO method. Moreover, it has been shown that the scaling of GTEMPO is better on the real-time axis than on the imaginary-time axis~\cite{ChenGuo2024b,ChenGuo2024c}, since the anti-periodic boundary condition in the latter case would result in a larger bond dimension of the underlying GMPS. Due to the close relationship of the multi-flavor GTEMPO method to GTEMPO, we believe that it could be even more powerful for the (non-equilibrium) real-time evolution.

\gcc{The data related to this paper is archived at~\cite{DataRepo}.}

\begin{acknowledgments}
Z. L. is partially supported by NSFC (22393913) and the robotic AI-Scientist platform of Chinese Academy of Sciences.
R. C. acknowledges support from National Natural Science Foundation of China under Grant No. 12104328. 
\end{acknowledgments}

\appendix

\section{More details of the multi-flavor GTEMPO method}\label{app:details}

\gcc{The GTEMPO method is a fermionic analogy of the TEMPO method for bosonic impurity problems. 
In TEMPO, a normal MPS is used to represent the integrand of the discretized bosonic PI, e.g., the ADT, as it is a tensor of normal scalars.
While in GTEMPO, we use GMPS to represent the ADT for fermionic PI, as the latter is a tensor of GVs (Grassmann tensor).
The central ideal of GTEMPO to solve quantum impurity problems is to represent the ADT as a GMPS and then compute multi-time impurity correlations based on it. Implementation-wise, this is achieved in three major steps: (1) Discretization of the fermionic PI in Eq.(\ref{eq:Z}), this is done by linearly discretizing the continuous Grassmann trajectories $\bolda_p(\tau)$, $\boldabar_p(\tau)$ into discrete GVs $\{a_{p,j}\}$ and $\{\abar_{p,j}\}$ where $j$ labels the discrete time step, after which the ADT naturally becomes a Grassmann tensor; (2) Constructing the bare impurity dynamics part $\gK[\boldabar\bolda]$ and each Feynman-Vernon IF $\gI_p[\boldabar_p\bolda_p]$ as GMPSs; and (3) Multiplying $\gK[\boldabar\bolda]$ and $\gI_p[\boldabar_p\bolda_p]$ together to obtain the ADT using Eq.(\ref{eq:adt}) and calculating multi-time impurity correlations based on it, e.g., Eq.(\ref{eq:Matsubara}).
For step (1), there is a well-established scheme for the discretization, referred to as the the quasi-adiabatic propagator path integral (QuAPI) method~\cite{makarov1994-path,makri1995-numerical}, for which one could also refer to Refs.~\cite{ChenGuo2024g,XuChen2024} for details. For step (2), essentially the only operation (besides the MPS bond truncation, which is a versatile operation that should be performed whenever the bond dimension of the MPS significantly grows) required is the multiplication between two GMPSs. For step (3), one needs to perform GMPS multiplication, as well as the integration of GMPS to obtain a scalar in the end. In addition, in step (3) we use the zipup algorithm to speed up the calculation of observables, which is a technique that evaluates the multiplication of several GMPSs on the fly instead of actually performing it, as a common practice in MPS algorithms to save memory and for computational efficiency.
The only additional operation of GMPS required in multi-flavor GTEMPO, on top of vanilla GTEMPO, is the partial integration of a GMPS, which returns another GMPS with less sites instead of a scalar.
}

\gcc{For make this work self-contained, we now discuss two key operations of GMPS, including GMPS multiplication, integration and partial integration of GMPS. 
We will mostly explain these operations based on their bosonic counterparts, i.e., the corresponding operations in TEMPO, as the latter are easier to understand, and then generalize to the fermionic case. One could also refer to Ref.~\cite{ChenGuo2024a,GuoChen2024d} for those details.
}

\gcc{The first important operation for manipulating the GMPS is the element-wise product. In case of normal tensor, this operation is simply 
\begin{align}
Z_{ijk\cdots} = X_{ijk\cdots}Y_{ijk\cdots},
\end{align}
for two tensors $X$, $Y$ as an example. 
This operation can be straightforwardly extended to normal MPSs with two steps: (1) One could first perform a tensor product between the two MPSs. Assuming that the site tensors of these two GMPSs are written as $X_{\alpha_{i-1},\alpha_i}^{\sigma_i}$ and $Y_{\beta_{i-1},\beta_i}^{\sigma_i^{\prime}}$ where $\sigma_i, \sigma_i'$ are physical indices and $\alpha_i, \beta_i$ are auxiliary indices, then the site tensor of the resulting MPS after the tensor product can be written as
\begin{align}
Z_{\alpha_{i-1}\beta_{i-1}, \alpha_i\beta_i}^{\sigma_i\sigma_i^{\prime}} = X_{\alpha_{i-1},\alpha_i}^{\sigma_i}Y_{\beta_{i-1},\beta_i}^{\sigma_i^{\prime}};
\end{align}
(2) Then one can perform the element-wise product for each pair of physical indices as
\begin{align}
W_{\alpha_{i-1}\beta_{i-1}, \alpha_i\beta_i}^{\sigma_i} = Z_{\alpha_{i-1}\beta_{i-1}, \alpha_i\beta_i}^{\sigma_i\sigma_i},
\end{align}
with $W_{\alpha_{i-1}\beta_{i-1}, \alpha_i\beta_i}^{\sigma_i}$ the site tensor of the final resulting MPS.
We can see that the bond dimension of the resulting MPS is exactly the product of the bond dimensions of the two input MPSs, if MPS bond truncation is not performed.
For two rank-$3$ Grassmann tensors denoted as $\mathcal{X} = \sum_{ijk}X_{ijk}\xi_1^{i}\xi_2^{j}\xi_3^{k}$ and $\mathcal{Y} = \sum_{ijk}Y_{ijk}\xi_1^{i}\xi_2^{j}\xi_3^{k}$, where $X$, $Y$ are coefficient tensors, and $\xi_l$ denotes the GV, the ``element-wise'' product is defined as
\begin{align}
\mathcal{Z} &= \mathcal{X}\mathcal{Y} = \sum_{ijk}X_{ijk}\xi_1^{i}\xi_2^{j}\xi_3^{k} \sum_{i'j'k'}Y_{i'j'k'}\xi_1^{i'}\xi_2^{j'}\xi_3^{k'} \nonumber \\
&=\sum_{ijki'j'k'} (-1)^{2i'+k+j} (-1)^{j'+k} X_{ijk}Y_{i'j'k'} \xi_1^{i+i'}\xi_2^{j+j'} \xi_3^{k+k'},
\end{align}
where the two sign factors $(-1)^{2i'+k+j}$ and $(-1)^{j'+k}$ are due to the change of locations of the two GVs $\xi_1^{i'}$ and $\xi_1^{j'}$. The result of the element-wise product is still a rank-$3$ Grassmann tensor by recognizing that $\xi^0 = 1, \xi^1=\xi, \xi^2 = 0$ from the definition of GV. The GMPS multiplication inherits from the ``element-wise'' product between two Grassmann tensors, similar to the bosonic case, and the detailed implementation can be found in Ref.~\cite{ChenGuo2024a}.
}

\gcc{The second important operation for manipulating the GMPS is the integration and partial integration. To illustrate this operation we consider the integration of a pair of GVs, $\xi_1$ and $\xi_2$, from a rank-$3$ Grassmann tensor $\mathcal{X} = \sum_{ijk}X_{ijk}\xi_1^{i}\xi_2^{j}\xi_3^{k}$. This operation can be performed as
\begin{align}
&\int\mathcal{D}[\xi_1\xi_2]\sum_{ijk}X_{ijk}\xi_1^{i}\xi_2^{j}\xi_3^{k} \nonumber \\ 
=& \int \dd\xi_1\dd\xi_2e^{-\xi_1\xi_2} \sum_{ijk}X_{ijk}\xi_1^{i}\xi_2^{j}\xi_3^{k} \nonumber \\ 
=& \sum_k\left(\sum_i X_{iik}\right) \xi_3^{k},
\end{align}
which results in a rank-$1$ Grassmann tensor in the end. In actual implementation of the Grassmann tensors, one only needs to stores the information of the coefficient tensor, and we can see that to perform a partial integration of the first pair of GVs one only needs to perform a partial trace of the first two indices of the coefficient tensor. In the general case that the pair of GVs been integrated out are not next to each other, one needs to first swap the GVs to put them in nearby positions, during which some sign factors may occur. 
Similarly, if we denote two nearby site tensors of a GMPS as $\sum_{\bar{\alpha}_{k+1}, \sigma_{k+1},\alpha_{k}}A^{\sigma_{k+1}}_{\bar{\alpha}_{k+1}, \alpha_{k}}\etabar_{k+1}^{\bar{\alpha}_{k+1}}\xi_{k+1}^{\sigma_{k+1}}\eta^{\alpha_{k}}$ and $\sum_{\bar{\alpha}_{k}, \sigma_{k},\alpha_{k-1}}A^{\sigma_{k}}_{\bar{\alpha}_{k}, \alpha_{k-1}} \etabar_{k}^{\bar{\alpha}_{k}}\xi_{k}^{\sigma_{k}}\eta^{\alpha_{k-1}}$, where $\bar{\eta}_k, \eta_k$ denote the auxiliary GVs and $\xi_k$ denotes the physical GV, 
then we could integrate out these two physical GVs as follows:
\begin{align}
&\int \dd\xi_k\dd\xi_{k+1}e^{-\xi_k\xi_{k+1}} \int \dd\etabar_k\dd \eta_k e^{-\etabar_k\eta_k} \nonumber \\ 
&\left(\sum_{\sigma_{k+1}, \bar{\alpha}_{k+1}, \alpha_k } A^{\sigma_{k+1}}_{\bar{\alpha}_{k+1}, \alpha_k} \etabar_{k+1}^{\bar{\alpha}_{k+1}} \xi_{k+1}^{\sigma_{k+1}} \eta_{k}^{\alpha_{k}}\right) \times \nonumber \\ &\left(\sum_{\sigma_{k}, \bar{\alpha}_{k}, \alpha_{k-1} } A^{\sigma_{k}}_{\bar{\alpha}_{k}, \alpha_{k-1}}  \etabar_{k}^{\bar{\alpha}_{k}}\xi_{k}^{\sigma_k} \eta_{k-1}^{\alpha_{k-1}}\right) \nonumber \\ 
=& \sum_{\bar{\alpha}_{k+1}, \alpha_{k-1}} \left(\sum_{\sigma_k, \alpha_k}  A^{\sigma_{k}}_{\bar{\alpha}_{k+1}, \alpha_k} A^{\sigma_{k}}_{\alpha_{k}, \alpha_{k-1}} \right) \etabar_{k+1}^{\bar{\alpha}_{k+1}} \eta_{k-1}^{\alpha_{k-1}},
\end{align}
which boils down to the normal contraction of the two coefficient tensors.
The result is a rank-$2$ Grassmann tensor without physical GVs, which can then be absorbed into nearby site tensors, and the net result is a new GMPS with the two sites corresponding to $\xi_{k+1}$ and $\xi_k$ missing compared to the original one. 
The partial integration, performed after the GMPS multiplication, is the major difference between our method to calculate the reduced ADT and the boundary MPS method to compress a 2D tensor network into 1D. As the information of the unused flavors are forgotten during this process, we observe that the bond dimension of resulting GMPS, i.e., $\chi_2$ can be greatly suppressed and grow much slower than the worst case $O(2^n\chi^n)$, which is as expected.
For full integration, one simply integrates out all the pairs of physical GVs, which would result in a single scalar.
}


\begin{thebibliography}{87}%
\makeatletter
\providecommand \@ifxundefined [1]{%
 \@ifx{#1\undefined}
}%
\providecommand \@ifnum [1]{%
 \ifnum #1\expandafter \@firstoftwo
 \else \expandafter \@secondoftwo
 \fi
}%
\providecommand \@ifx [1]{%
 \ifx #1\expandafter \@firstoftwo
 \else \expandafter \@secondoftwo
 \fi
}%
\providecommand \natexlab [1]{#1}%
\providecommand \enquote  [1]{``#1''}%
\providecommand \bibnamefont  [1]{#1}%
\providecommand \bibfnamefont [1]{#1}%
\providecommand \citenamefont [1]{#1}%
\providecommand \href@noop [0]{\@secondoftwo}%
\providecommand \href [0]{\begingroup \@sanitize@url \@href}%
\providecommand \@href[1]{\@@startlink{#1}\@@href}%
\providecommand \@@href[1]{\endgroup#1\@@endlink}%
\providecommand \@sanitize@url [0]{\catcode `\\12\catcode `\$12\catcode
  `\&12\catcode `\#12\catcode `\^12\catcode `\_12\catcode `\%12\relax}%
\providecommand \@@startlink[1]{}%
\providecommand \@@endlink[0]{}%
\providecommand \url  [0]{\begingroup\@sanitize@url \@url }%
\providecommand \@url [1]{\endgroup\@href {#1}{\urlprefix }}%
\providecommand \urlprefix  [0]{URL }%
\providecommand \Eprint [0]{\href }%
\providecommand \doibase [0]{https://doi.org/}%
\providecommand \selectlanguage [0]{\@gobble}%
\providecommand \bibinfo  [0]{\@secondoftwo}%
\providecommand \bibfield  [0]{\@secondoftwo}%
\providecommand \translation [1]{[#1]}%
\providecommand \BibitemOpen [0]{}%
\providecommand \bibitemStop [0]{}%
\providecommand \bibitemNoStop [0]{.\EOS\space}%
\providecommand \EOS [0]{\spacefactor3000\relax}%
\providecommand \BibitemShut  [1]{\csname bibitem#1\endcsname}%
\let\auto@bib@innerbib\@empty
\bibitem [{\citenamefont {Georges}\ \emph
  {et~al.}(1996{\natexlab{a}})\citenamefont {Georges}, \citenamefont {Kotliar},
  \citenamefont {Krauth},\ and\ \citenamefont
  {Rozenberg}}]{GeorgesRozenberg1996}%
  \BibitemOpen
  \bibfield  {author} {\bibinfo {author} {\bibfnamefont {A.}~\bibnamefont
  {Georges}}, \bibinfo {author} {\bibfnamefont {G.}~\bibnamefont {Kotliar}},
  \bibinfo {author} {\bibfnamefont {W.}~\bibnamefont {Krauth}},\ and\ \bibinfo
  {author} {\bibfnamefont {M.~J.}\ \bibnamefont {Rozenberg}},\ }\bibfield
  {title} {\bibinfo {title} {Dynamical mean-field theory of strongly correlated
  fermion systems and the limit of infinite dimensions},\ }\href
  {https://doi.org/10.1103/RevModPhys.68.13} {\bibfield  {journal} {\bibinfo
  {journal} {Rev. Mod. Phys.}\ }\textbf {\bibinfo {volume} {68}},\ \bibinfo
  {pages} {13} (\bibinfo {year} {1996}{\natexlab{a}})}\BibitemShut {NoStop}%
\bibitem [{\citenamefont {Georges}\ and\ \citenamefont
  {Kotliar}(1992)}]{GeorgesKotliar1992}%
  \BibitemOpen
  \bibfield  {author} {\bibinfo {author} {\bibfnamefont {A.}~\bibnamefont
  {Georges}}\ and\ \bibinfo {author} {\bibfnamefont {G.}~\bibnamefont
  {Kotliar}},\ }\bibfield  {title} {\bibinfo {title} {Hubbard model in infinite
  dimensions},\ }\href {https://doi.org/10.1103/PhysRevB.45.6479} {\bibfield
  {journal} {\bibinfo  {journal} {Phys. Rev. B}\ }\textbf {\bibinfo {volume}
  {45}},\ \bibinfo {pages} {6479} (\bibinfo {year} {1992})}\BibitemShut
  {NoStop}%
\bibitem [{\citenamefont {Metzner}\ and\ \citenamefont
  {Vollhardt}(1989)}]{MetznerVollhardt1989}%
  \BibitemOpen
  \bibfield  {author} {\bibinfo {author} {\bibfnamefont {W.}~\bibnamefont
  {Metzner}}\ and\ \bibinfo {author} {\bibfnamefont {D.}~\bibnamefont
  {Vollhardt}},\ }\bibfield  {title} {\bibinfo {title} {Correlated lattice
  fermions in $d=\ensuremath{\infty}$ dimensions},\ }\href
  {https://doi.org/10.1103/PhysRevLett.62.324} {\bibfield  {journal} {\bibinfo
  {journal} {Phys. Rev. Lett.}\ }\textbf {\bibinfo {volume} {62}},\ \bibinfo
  {pages} {324} (\bibinfo {year} {1989})}\BibitemShut {NoStop}%
\bibitem [{\citenamefont {Kotliar}\ \emph {et~al.}(2006)\citenamefont
  {Kotliar}, \citenamefont {Savrasov}, \citenamefont {Haule}, \citenamefont
  {Oudovenko}, \citenamefont {Parcollet},\ and\ \citenamefont
  {Marianetti}}]{KotliarMarianetti2006}%
  \BibitemOpen
  \bibfield  {author} {\bibinfo {author} {\bibfnamefont {G.}~\bibnamefont
  {Kotliar}}, \bibinfo {author} {\bibfnamefont {S.~Y.}\ \bibnamefont
  {Savrasov}}, \bibinfo {author} {\bibfnamefont {K.}~\bibnamefont {Haule}},
  \bibinfo {author} {\bibfnamefont {V.~S.}\ \bibnamefont {Oudovenko}}, \bibinfo
  {author} {\bibfnamefont {O.}~\bibnamefont {Parcollet}},\ and\ \bibinfo
  {author} {\bibfnamefont {C.~A.}\ \bibnamefont {Marianetti}},\ }\bibfield
  {title} {\bibinfo {title} {Electronic structure calculations with dynamical
  mean-field theory},\ }\href {https://doi.org/10.1103/RevModPhys.78.865}
  {\bibfield  {journal} {\bibinfo  {journal} {Rev. Mod. Phys.}\ }\textbf
  {\bibinfo {volume} {78}},\ \bibinfo {pages} {865} (\bibinfo {year}
  {2006})}\BibitemShut {NoStop}%
\bibitem [{\citenamefont {Rubtsov}\ and\ \citenamefont
  {Lichtenstein}(2004)}]{RubtsovLichtenstein2004}%
  \BibitemOpen
  \bibfield  {author} {\bibinfo {author} {\bibfnamefont {A.~N.}\ \bibnamefont
  {Rubtsov}}\ and\ \bibinfo {author} {\bibfnamefont {A.~I.}\ \bibnamefont
  {Lichtenstein}},\ }\bibfield  {title} {\bibinfo {title} {Continuous-time
  quantum monte carlo method for fermions: Beyond auxiliary field framework},\
  }\href {https://doi.org/10.1134/1.1800216} {\bibfield  {journal} {\bibinfo
  {journal} {J. Exp. Theor. Phys. Lett.}\ }\textbf {\bibinfo {volume} {80}},\
  \bibinfo {pages} {61} (\bibinfo {year} {2004})}\BibitemShut {NoStop}%
\bibitem [{\citenamefont {Rubtsov}\ \emph {et~al.}(2005)\citenamefont
  {Rubtsov}, \citenamefont {Savkin},\ and\ \citenamefont
  {Lichtenstein}}]{RubtsovLichtenstein2005}%
  \BibitemOpen
  \bibfield  {author} {\bibinfo {author} {\bibfnamefont {A.~N.}\ \bibnamefont
  {Rubtsov}}, \bibinfo {author} {\bibfnamefont {V.~V.}\ \bibnamefont
  {Savkin}},\ and\ \bibinfo {author} {\bibfnamefont {A.~I.}\ \bibnamefont
  {Lichtenstein}},\ }\bibfield  {title} {\bibinfo {title} {Continuous-time
  quantum monte carlo method for fermions},\ }\href
  {https://doi.org/10.1103/PhysRevB.72.035122} {\bibfield  {journal} {\bibinfo
  {journal} {Phys. Rev. B}\ }\textbf {\bibinfo {volume} {72}},\ \bibinfo
  {pages} {035122} (\bibinfo {year} {2005})}\BibitemShut {NoStop}%
\bibitem [{\citenamefont {Werner}\ \emph {et~al.}(2006)\citenamefont {Werner},
  \citenamefont {Comanac}, \citenamefont {de' Medici}, \citenamefont {Troyer},\
  and\ \citenamefont {Millis}}]{WernerMillis2006}%
  \BibitemOpen
  \bibfield  {author} {\bibinfo {author} {\bibfnamefont {P.}~\bibnamefont
  {Werner}}, \bibinfo {author} {\bibfnamefont {A.}~\bibnamefont {Comanac}},
  \bibinfo {author} {\bibfnamefont {L.}~\bibnamefont {de' Medici}}, \bibinfo
  {author} {\bibfnamefont {M.}~\bibnamefont {Troyer}},\ and\ \bibinfo {author}
  {\bibfnamefont {A.~J.}\ \bibnamefont {Millis}},\ }\bibfield  {title}
  {\bibinfo {title} {Continuous-time solver for quantum impurity models},\
  }\href {https://doi.org/10.1103/PhysRevLett.97.076405} {\bibfield  {journal}
  {\bibinfo  {journal} {Phys. Rev. Lett.}\ }\textbf {\bibinfo {volume} {97}},\
  \bibinfo {pages} {076405} (\bibinfo {year} {2006})}\BibitemShut {NoStop}%
\bibitem [{\citenamefont {Gull}\ \emph {et~al.}(2008)\citenamefont {Gull},
  \citenamefont {Werner}, \citenamefont {Parcollet},\ and\ \citenamefont
  {Troyer}}]{GullTroyer2008}%
  \BibitemOpen
  \bibfield  {author} {\bibinfo {author} {\bibfnamefont {E.}~\bibnamefont
  {Gull}}, \bibinfo {author} {\bibfnamefont {P.}~\bibnamefont {Werner}},
  \bibinfo {author} {\bibfnamefont {O.}~\bibnamefont {Parcollet}},\ and\
  \bibinfo {author} {\bibfnamefont {M.}~\bibnamefont {Troyer}},\ }\bibfield
  {title} {\bibinfo {title} {Continuous-time auxiliary-field monte carlo for
  quantum impurity models},\ }\href
  {https://doi.org/10.1209/0295-5075/82/57003} {\bibfield  {journal} {\bibinfo
  {journal} {EPL}\ }\textbf {\bibinfo {volume} {82}},\ \bibinfo {pages} {57003}
  (\bibinfo {year} {2008})}\BibitemShut {NoStop}%
\bibitem [{\citenamefont {Chan}\ \emph {et~al.}(2009)\citenamefont {Chan},
  \citenamefont {Werner},\ and\ \citenamefont {Millis}}]{ChanMillis2009}%
  \BibitemOpen
  \bibfield  {author} {\bibinfo {author} {\bibfnamefont {C.-K.}\ \bibnamefont
  {Chan}}, \bibinfo {author} {\bibfnamefont {P.}~\bibnamefont {Werner}},\ and\
  \bibinfo {author} {\bibfnamefont {A.~J.}\ \bibnamefont {Millis}},\ }\bibfield
   {title} {\bibinfo {title} {Magnetism and orbital ordering in an interacting
  three-band model: A dynamical mean-field study},\ }\href
  {https://doi.org/10.1103/PhysRevB.80.235114} {\bibfield  {journal} {\bibinfo
  {journal} {Phys. Rev. B}\ }\textbf {\bibinfo {volume} {80}},\ \bibinfo
  {pages} {235114} (\bibinfo {year} {2009})}\BibitemShut {NoStop}%
\bibitem [{\citenamefont {Haule}\ \emph {et~al.}(2010)\citenamefont {Haule},
  \citenamefont {Yee},\ and\ \citenamefont {Kim}}]{haule2010-dynamical}%
  \BibitemOpen
  \bibfield  {author} {\bibinfo {author} {\bibfnamefont {K.}~\bibnamefont
  {Haule}}, \bibinfo {author} {\bibfnamefont {C.-H.}\ \bibnamefont {Yee}},\
  and\ \bibinfo {author} {\bibfnamefont {K.}~\bibnamefont {Kim}},\ }\bibfield
  {title} {\bibinfo {title} {Dynamical mean-field theory within the
  full-potential methods: Electronic structure of {CeIrIn5}, {CeCoIn5}, and
  {CeRhIn5}},\ }\href {https://doi.org/10.1103/physrevb.81.195107} {\bibfield
  {journal} {\bibinfo  {journal} {Phys. Rev. B}\ }\textbf {\bibinfo {volume}
  {81}},\ \bibinfo {pages} {195107} (\bibinfo {year} {2010})}\BibitemShut
  {NoStop}%
\bibitem [{\citenamefont {Gull}\ \emph {et~al.}(2011)\citenamefont {Gull},
  \citenamefont {Millis}, \citenamefont {Lichtenstein}, \citenamefont
  {Rubtsov}, \citenamefont {Troyer},\ and\ \citenamefont
  {Werner}}]{GullWerner2011}%
  \BibitemOpen
  \bibfield  {author} {\bibinfo {author} {\bibfnamefont {E.}~\bibnamefont
  {Gull}}, \bibinfo {author} {\bibfnamefont {A.~J.}\ \bibnamefont {Millis}},
  \bibinfo {author} {\bibfnamefont {A.~I.}\ \bibnamefont {Lichtenstein}},
  \bibinfo {author} {\bibfnamefont {A.~N.}\ \bibnamefont {Rubtsov}}, \bibinfo
  {author} {\bibfnamefont {M.}~\bibnamefont {Troyer}},\ and\ \bibinfo {author}
  {\bibfnamefont {P.}~\bibnamefont {Werner}},\ }\bibfield  {title} {\bibinfo
  {title} {Continuous-time monte carlo methods for quantum impurity models},\
  }\href {https://doi.org/10.1103/RevModPhys.83.349} {\bibfield  {journal}
  {\bibinfo  {journal} {Rev. Mod. Phys.}\ }\textbf {\bibinfo {volume} {83}},\
  \bibinfo {pages} {349} (\bibinfo {year} {2011})}\BibitemShut {NoStop}%
\bibitem [{\citenamefont {Huang}\ \emph {et~al.}(2014)\citenamefont {Huang},
  \citenamefont {Wehling},\ and\ \citenamefont
  {Werner}}]{huang2014-electronic}%
  \BibitemOpen
  \bibfield  {author} {\bibinfo {author} {\bibfnamefont {L.}~\bibnamefont
  {Huang}}, \bibinfo {author} {\bibfnamefont {T.~O.}\ \bibnamefont {Wehling}},\
  and\ \bibinfo {author} {\bibfnamefont {P.}~\bibnamefont {Werner}},\
  }\bibfield  {title} {\bibinfo {title} {Electronic excitation spectra of the
  five-orbital anderson impurity model: From the atomic limit to itinerant
  atomic magnetism},\ }\href {https://doi.org/10.1103/physrevb.89.245104}
  {\bibfield  {journal} {\bibinfo  {journal} {Phys. Rev. B}\ }\textbf {\bibinfo
  {volume} {89}},\ \bibinfo {pages} {245104} (\bibinfo {year}
  {2014})}\BibitemShut {NoStop}%
\bibitem [{\citenamefont {Lu}\ and\ \citenamefont
  {Huang}(2016)}]{lu2016-pressure}%
  \BibitemOpen
  \bibfield  {author} {\bibinfo {author} {\bibfnamefont {H.}~\bibnamefont
  {Lu}}\ and\ \bibinfo {author} {\bibfnamefont {L.}~\bibnamefont {Huang}},\
  }\bibfield  {title} {\bibinfo {title} {Pressure-driven4flocalized-itinerant
  crossover in heavy-fermion compoundcein3: A first-principles many-body
  perspective},\ }\href {https://doi.org/10.1103/physrevb.94.075132} {\bibfield
   {journal} {\bibinfo  {journal} {Phys. Rev. B}\ }\textbf {\bibinfo {volume}
  {94}},\ \bibinfo {pages} {075132} (\bibinfo {year} {2016})}\BibitemShut
  {NoStop}%
\bibitem [{\citenamefont {Yue}\ and\ \citenamefont
  {Werner}(2021)}]{yue2021-pairing}%
  \BibitemOpen
  \bibfield  {author} {\bibinfo {author} {\bibfnamefont {C.}~\bibnamefont
  {Yue}}\ and\ \bibinfo {author} {\bibfnamefont {P.}~\bibnamefont {Werner}},\
  }\bibfield  {title} {\bibinfo {title} {Pairing enhanced by local orbital
  fluctuations in a model for monolayer fese},\ }\href
  {https://doi.org/10.1103/physrevb.104.184507} {\bibfield  {journal} {\bibinfo
   {journal} {Phys. Rev. B}\ }\textbf {\bibinfo {volume} {104}},\ \bibinfo
  {pages} {184507} (\bibinfo {year} {2021})}\BibitemShut {NoStop}%
\bibitem [{\citenamefont {Caffarel}\ and\ \citenamefont
  {Krauth}(1994)}]{CaffarelKrauth1994}%
  \BibitemOpen
  \bibfield  {author} {\bibinfo {author} {\bibfnamefont {M.}~\bibnamefont
  {Caffarel}}\ and\ \bibinfo {author} {\bibfnamefont {W.}~\bibnamefont
  {Krauth}},\ }\bibfield  {title} {\bibinfo {title} {Exact diagonalization
  approach to correlated fermions in infinite dimensions: Mott transition and
  superconductivity},\ }\href {https://doi.org/10.1103/PhysRevLett.72.1545}
  {\bibfield  {journal} {\bibinfo  {journal} {Phys. Rev. Lett.}\ }\textbf
  {\bibinfo {volume} {72}},\ \bibinfo {pages} {1545} (\bibinfo {year}
  {1994})}\BibitemShut {NoStop}%
\bibitem [{\citenamefont {Koch}\ \emph {et~al.}(2008)\citenamefont {Koch},
  \citenamefont {Sangiovanni},\ and\ \citenamefont
  {Gunnarsson}}]{KochGunnarsson2008}%
  \BibitemOpen
  \bibfield  {author} {\bibinfo {author} {\bibfnamefont {E.}~\bibnamefont
  {Koch}}, \bibinfo {author} {\bibfnamefont {G.}~\bibnamefont {Sangiovanni}},\
  and\ \bibinfo {author} {\bibfnamefont {O.}~\bibnamefont {Gunnarsson}},\
  }\bibfield  {title} {\bibinfo {title} {Sum rules and bath parametrization for
  quantum cluster theories},\ }\href
  {https://doi.org/10.1103/PhysRevB.78.115102} {\bibfield  {journal} {\bibinfo
  {journal} {Phys. Rev. B}\ }\textbf {\bibinfo {volume} {78}},\ \bibinfo
  {pages} {115102} (\bibinfo {year} {2008})}\BibitemShut {NoStop}%
\bibitem [{\citenamefont {Granath}\ and\ \citenamefont
  {Strand}(2012)}]{GranathStrand2012}%
  \BibitemOpen
  \bibfield  {author} {\bibinfo {author} {\bibfnamefont {M.}~\bibnamefont
  {Granath}}\ and\ \bibinfo {author} {\bibfnamefont {H.~U.~R.}\ \bibnamefont
  {Strand}},\ }\bibfield  {title} {\bibinfo {title} {Distributional exact
  diagonalization formalism for quantum impurity models},\ }\href
  {https://doi.org/10.1103/PhysRevB.86.115111} {\bibfield  {journal} {\bibinfo
  {journal} {Phys. Rev. B}\ }\textbf {\bibinfo {volume} {86}},\ \bibinfo
  {pages} {115111} (\bibinfo {year} {2012})}\BibitemShut {NoStop}%
\bibitem [{\citenamefont {Lu}\ \emph {et~al.}(2014)\citenamefont {Lu},
  \citenamefont {H\"oppner}, \citenamefont {Gunnarsson},\ and\ \citenamefont
  {Haverkort}}]{LuHaverkort2014}%
  \BibitemOpen
  \bibfield  {author} {\bibinfo {author} {\bibfnamefont {Y.}~\bibnamefont
  {Lu}}, \bibinfo {author} {\bibfnamefont {M.}~\bibnamefont {H\"oppner}},
  \bibinfo {author} {\bibfnamefont {O.}~\bibnamefont {Gunnarsson}},\ and\
  \bibinfo {author} {\bibfnamefont {M.~W.}\ \bibnamefont {Haverkort}},\
  }\bibfield  {title} {\bibinfo {title} {Efficient real-frequency solver for
  dynamical mean-field theory},\ }\href
  {https://doi.org/10.1103/PhysRevB.90.085102} {\bibfield  {journal} {\bibinfo
  {journal} {Phys. Rev. B}\ }\textbf {\bibinfo {volume} {90}},\ \bibinfo
  {pages} {085102} (\bibinfo {year} {2014})}\BibitemShut {NoStop}%
\bibitem [{\citenamefont {Mejuto-Zaera}\ \emph {et~al.}(2020)\citenamefont
  {Mejuto-Zaera}, \citenamefont {Zepeda-N\'u\~nez}, \citenamefont {Lindsey},
  \citenamefont {Tubman}, \citenamefont {Whaley},\ and\ \citenamefont
  {Lin}}]{ZaeraLin2020}%
  \BibitemOpen
  \bibfield  {author} {\bibinfo {author} {\bibfnamefont {C.}~\bibnamefont
  {Mejuto-Zaera}}, \bibinfo {author} {\bibfnamefont {L.}~\bibnamefont
  {Zepeda-N\'u\~nez}}, \bibinfo {author} {\bibfnamefont {M.}~\bibnamefont
  {Lindsey}}, \bibinfo {author} {\bibfnamefont {N.}~\bibnamefont {Tubman}},
  \bibinfo {author} {\bibfnamefont {B.}~\bibnamefont {Whaley}},\ and\ \bibinfo
  {author} {\bibfnamefont {L.}~\bibnamefont {Lin}},\ }\bibfield  {title}
  {\bibinfo {title} {Efficient hybridization fitting for dynamical mean-field
  theory via semi-definite relaxation},\ }\href
  {https://doi.org/10.1103/PhysRevB.101.035143} {\bibfield  {journal} {\bibinfo
   {journal} {Phys. Rev. B}\ }\textbf {\bibinfo {volume} {101}},\ \bibinfo
  {pages} {035143} (\bibinfo {year} {2020})}\BibitemShut {NoStop}%
\bibitem [{\citenamefont {Lu}\ \emph {et~al.}(2019)\citenamefont {Lu},
  \citenamefont {Cao}, \citenamefont {Hansmann},\ and\ \citenamefont
  {Haverkort}}]{LuHaverkort2019}%
  \BibitemOpen
  \bibfield  {author} {\bibinfo {author} {\bibfnamefont {Y.}~\bibnamefont
  {Lu}}, \bibinfo {author} {\bibfnamefont {X.}~\bibnamefont {Cao}}, \bibinfo
  {author} {\bibfnamefont {P.}~\bibnamefont {Hansmann}},\ and\ \bibinfo
  {author} {\bibfnamefont {M.~W.}\ \bibnamefont {Haverkort}},\ }\bibfield
  {title} {\bibinfo {title} {Natural-orbital impurity solver and projection
  approach for green's functions},\ }\href
  {https://doi.org/10.1103/PhysRevB.100.115134} {\bibfield  {journal} {\bibinfo
   {journal} {Phys. Rev. B}\ }\textbf {\bibinfo {volume} {100}},\ \bibinfo
  {pages} {115134} (\bibinfo {year} {2019})}\BibitemShut {NoStop}%
\bibitem [{\citenamefont {He}\ and\ \citenamefont {Lu}(2014)}]{HeLu2014}%
  \BibitemOpen
  \bibfield  {author} {\bibinfo {author} {\bibfnamefont {R.-Q.}\ \bibnamefont
  {He}}\ and\ \bibinfo {author} {\bibfnamefont {Z.-Y.}\ \bibnamefont {Lu}},\
  }\bibfield  {title} {\bibinfo {title} {Quantum renormalization groups based
  on natural orbitals},\ }\href {https://doi.org/10.1103/PhysRevB.89.085108}
  {\bibfield  {journal} {\bibinfo  {journal} {Phys. Rev. B}\ }\textbf {\bibinfo
  {volume} {89}},\ \bibinfo {pages} {085108} (\bibinfo {year}
  {2014})}\BibitemShut {NoStop}%
\bibitem [{\citenamefont {He}\ \emph {et~al.}(2015)\citenamefont {He},
  \citenamefont {Dai},\ and\ \citenamefont {Lu}}]{HeLu2015}%
  \BibitemOpen
  \bibfield  {author} {\bibinfo {author} {\bibfnamefont {R.-Q.}\ \bibnamefont
  {He}}, \bibinfo {author} {\bibfnamefont {J.}~\bibnamefont {Dai}},\ and\
  \bibinfo {author} {\bibfnamefont {Z.-Y.}\ \bibnamefont {Lu}},\ }\bibfield
  {title} {\bibinfo {title} {Natural orbitals renormalization group approach to
  the two-impurity kondo critical point},\ }\href
  {https://doi.org/10.1103/PhysRevB.91.155140} {\bibfield  {journal} {\bibinfo
  {journal} {Phys. Rev. B}\ }\textbf {\bibinfo {volume} {91}},\ \bibinfo
  {pages} {155140} (\bibinfo {year} {2015})}\BibitemShut {NoStop}%
\bibitem [{\citenamefont {Wilson}(1975)}]{Wilson1975}%
  \BibitemOpen
  \bibfield  {author} {\bibinfo {author} {\bibfnamefont {K.~G.}\ \bibnamefont
  {Wilson}},\ }\bibfield  {title} {\bibinfo {title} {The renormalization group:
  Critical phenomena and the kondo problem},\ }\href
  {https://doi.org/10.1103/RevModPhys.47.773} {\bibfield  {journal} {\bibinfo
  {journal} {Rev. Mod. Phys.}\ }\textbf {\bibinfo {volume} {47}},\ \bibinfo
  {pages} {773} (\bibinfo {year} {1975})}\BibitemShut {NoStop}%
\bibitem [{\citenamefont {Bulla}(1999)}]{Bulla1999}%
  \BibitemOpen
  \bibfield  {author} {\bibinfo {author} {\bibfnamefont {R.}~\bibnamefont
  {Bulla}},\ }\bibfield  {title} {\bibinfo {title} {Zero temperature
  metal-insulator transition in the infinite-dimensional hubbard model},\
  }\href {https://doi.org/10.1103/PhysRevLett.83.136} {\bibfield  {journal}
  {\bibinfo  {journal} {Phys. Rev. Lett.}\ }\textbf {\bibinfo {volume} {83}},\
  \bibinfo {pages} {136} (\bibinfo {year} {1999})}\BibitemShut {NoStop}%
\bibitem [{\citenamefont {Bulla}\ \emph {et~al.}(2008)\citenamefont {Bulla},
  \citenamefont {Costi},\ and\ \citenamefont {Pruschke}}]{BullaPruschke2008}%
  \BibitemOpen
  \bibfield  {author} {\bibinfo {author} {\bibfnamefont {R.}~\bibnamefont
  {Bulla}}, \bibinfo {author} {\bibfnamefont {T.~A.}\ \bibnamefont {Costi}},\
  and\ \bibinfo {author} {\bibfnamefont {T.}~\bibnamefont {Pruschke}},\
  }\bibfield  {title} {\bibinfo {title} {Numerical renormalization group method
  for quantum impurity systems},\ }\href
  {https://doi.org/10.1103/RevModPhys.80.395} {\bibfield  {journal} {\bibinfo
  {journal} {Rev. Mod. Phys.}\ }\textbf {\bibinfo {volume} {80}},\ \bibinfo
  {pages} {395} (\bibinfo {year} {2008})}\BibitemShut {NoStop}%
\bibitem [{\citenamefont {Anders}(2008)}]{Frithjof2008}%
  \BibitemOpen
  \bibfield  {author} {\bibinfo {author} {\bibfnamefont {F.~B.}\ \bibnamefont
  {Anders}},\ }\bibfield  {title} {\bibinfo {title} {A numerical
  renormalization group approach to non-equilibrium green functions for quantum
  impurity models},\ }\href {https://doi.org/10.1088/0953-8984/20/19/195216}
  {\bibfield  {journal} {\bibinfo  {journal} {J. Phys. Condens. Matter}\
  }\textbf {\bibinfo {volume} {20}},\ \bibinfo {pages} {195216} (\bibinfo
  {year} {2008})}\BibitemShut {NoStop}%
\bibitem [{\citenamefont {\ifmmode~\check{Z}\else \v{Z}\fi{}itko}\ and\
  \citenamefont {Pruschke}(2009)}]{ZitkoPruschke2009}%
  \BibitemOpen
  \bibfield  {author} {\bibinfo {author} {\bibfnamefont {R.}~\bibnamefont
  {\ifmmode~\check{Z}\else \v{Z}\fi{}itko}}\ and\ \bibinfo {author}
  {\bibfnamefont {T.}~\bibnamefont {Pruschke}},\ }\bibfield  {title} {\bibinfo
  {title} {Energy resolution and discretization artifacts in the numerical
  renormalization group},\ }\href {https://doi.org/10.1103/PhysRevB.79.085106}
  {\bibfield  {journal} {\bibinfo  {journal} {Phys. Rev. B}\ }\textbf {\bibinfo
  {volume} {79}},\ \bibinfo {pages} {085106} (\bibinfo {year}
  {2009})}\BibitemShut {NoStop}%
\bibitem [{\citenamefont {Deng}\ \emph {et~al.}(2013)\citenamefont {Deng},
  \citenamefont {Mravlje}, \citenamefont {\ifmmode~\check{Z}\else
  \v{Z}\fi{}itko}, \citenamefont {Ferrero}, \citenamefont {Kotliar},\ and\
  \citenamefont {Georges}}]{DengGeorges2013}%
  \BibitemOpen
  \bibfield  {author} {\bibinfo {author} {\bibfnamefont {X.}~\bibnamefont
  {Deng}}, \bibinfo {author} {\bibfnamefont {J.}~\bibnamefont {Mravlje}},
  \bibinfo {author} {\bibfnamefont {R.}~\bibnamefont {\ifmmode~\check{Z}\else
  \v{Z}\fi{}itko}}, \bibinfo {author} {\bibfnamefont {M.}~\bibnamefont
  {Ferrero}}, \bibinfo {author} {\bibfnamefont {G.}~\bibnamefont {Kotliar}},\
  and\ \bibinfo {author} {\bibfnamefont {A.}~\bibnamefont {Georges}},\
  }\bibfield  {title} {\bibinfo {title} {How bad metals turn good:
  Spectroscopic signatures of resilient quasiparticles},\ }\href
  {https://doi.org/10.1103/PhysRevLett.110.086401} {\bibfield  {journal}
  {\bibinfo  {journal} {Phys. Rev. Lett.}\ }\textbf {\bibinfo {volume} {110}},\
  \bibinfo {pages} {086401} (\bibinfo {year} {2013})}\BibitemShut {NoStop}%
\bibitem [{\citenamefont {Stadler}\ \emph {et~al.}(2015)\citenamefont
  {Stadler}, \citenamefont {Yin}, \citenamefont {von Delft}, \citenamefont
  {Kotliar},\ and\ \citenamefont {Weichselbaum}}]{StadlerWeichselbaum2015}%
  \BibitemOpen
  \bibfield  {author} {\bibinfo {author} {\bibfnamefont {K.~M.}\ \bibnamefont
  {Stadler}}, \bibinfo {author} {\bibfnamefont {Z.~P.}\ \bibnamefont {Yin}},
  \bibinfo {author} {\bibfnamefont {J.}~\bibnamefont {von Delft}}, \bibinfo
  {author} {\bibfnamefont {G.}~\bibnamefont {Kotliar}},\ and\ \bibinfo {author}
  {\bibfnamefont {A.}~\bibnamefont {Weichselbaum}},\ }\bibfield  {title}
  {\bibinfo {title} {Dynamical mean-field theory plus numerical
  renormalization-group study of spin-orbital separation in a three-band hund
  metal},\ }\href {https://doi.org/10.1103/PhysRevLett.115.136401} {\bibfield
  {journal} {\bibinfo  {journal} {Phys. Rev. Lett.}\ }\textbf {\bibinfo
  {volume} {115}},\ \bibinfo {pages} {136401} (\bibinfo {year}
  {2015})}\BibitemShut {NoStop}%
\bibitem [{\citenamefont {Lee}\ and\ \citenamefont
  {Weichselbaum}(2016)}]{LeeWeichselbaum2016}%
  \BibitemOpen
  \bibfield  {author} {\bibinfo {author} {\bibfnamefont {S.-S.~B.}\
  \bibnamefont {Lee}}\ and\ \bibinfo {author} {\bibfnamefont {A.}~\bibnamefont
  {Weichselbaum}},\ }\bibfield  {title} {\bibinfo {title} {Adaptive broadening
  to improve spectral resolution in the numerical renormalization group},\
  }\href {https://doi.org/10.1103/PhysRevB.94.235127} {\bibfield  {journal}
  {\bibinfo  {journal} {Phys. Rev. B}\ }\textbf {\bibinfo {volume} {94}},\
  \bibinfo {pages} {235127} (\bibinfo {year} {2016})}\BibitemShut {NoStop}%
\bibitem [{\citenamefont {Lee}\ \emph {et~al.}(2017)\citenamefont {Lee},
  \citenamefont {von Delft},\ and\ \citenamefont
  {Weichselbaum}}]{LeeWeichselbaum2017}%
  \BibitemOpen
  \bibfield  {author} {\bibinfo {author} {\bibfnamefont {S.-S.~B.}\
  \bibnamefont {Lee}}, \bibinfo {author} {\bibfnamefont {J.}~\bibnamefont {von
  Delft}},\ and\ \bibinfo {author} {\bibfnamefont {A.}~\bibnamefont
  {Weichselbaum}},\ }\bibfield  {title} {\bibinfo {title} {Doublon-holon origin
  of the subpeaks at the hubbard band edges},\ }\href
  {https://doi.org/10.1103/PhysRevLett.119.236402} {\bibfield  {journal}
  {\bibinfo  {journal} {Phys. Rev. Lett.}\ }\textbf {\bibinfo {volume} {119}},\
  \bibinfo {pages} {236402} (\bibinfo {year} {2017})}\BibitemShut {NoStop}%
\bibitem [{\citenamefont {Cornaglia}\ \emph {et~al.}(2004)\citenamefont
  {Cornaglia}, \citenamefont {Ness},\ and\ \citenamefont
  {Grempel}}]{CornagliaGrempel2004}%
  \BibitemOpen
  \bibfield  {author} {\bibinfo {author} {\bibfnamefont {P.~S.}\ \bibnamefont
  {Cornaglia}}, \bibinfo {author} {\bibfnamefont {H.}~\bibnamefont {Ness}},\
  and\ \bibinfo {author} {\bibfnamefont {D.~R.}\ \bibnamefont {Grempel}},\
  }\bibfield  {title} {\bibinfo {title} {Many-body effects on the transport
  properties of single-molecule devices},\ }\href
  {https://doi.org/10.1103/PhysRevLett.93.147201} {\bibfield  {journal}
  {\bibinfo  {journal} {Phys. Rev. Lett.}\ }\textbf {\bibinfo {volume} {93}},\
  \bibinfo {pages} {147201} (\bibinfo {year} {2004})}\BibitemShut {NoStop}%
\bibitem [{\citenamefont {Paaske}\ and\ \citenamefont
  {Flensberg}(2005)}]{PaaskeFlensberg2005}%
  \BibitemOpen
  \bibfield  {author} {\bibinfo {author} {\bibfnamefont {J.}~\bibnamefont
  {Paaske}}\ and\ \bibinfo {author} {\bibfnamefont {K.}~\bibnamefont
  {Flensberg}},\ }\bibfield  {title} {\bibinfo {title} {Vibrational sidebands
  and the kondo effect in molecular transistors},\ }\href
  {https://doi.org/10.1103/PhysRevLett.94.176801} {\bibfield  {journal}
  {\bibinfo  {journal} {Phys. Rev. Lett.}\ }\textbf {\bibinfo {volume} {94}},\
  \bibinfo {pages} {176801} (\bibinfo {year} {2005})}\BibitemShut {NoStop}%
\bibitem [{\citenamefont {Cornaglia}\ \emph {et~al.}(2005)\citenamefont
  {Cornaglia}, \citenamefont {Grempel},\ and\ \citenamefont
  {Ness}}]{CornagliaNess2005}%
  \BibitemOpen
  \bibfield  {author} {\bibinfo {author} {\bibfnamefont {P.~S.}\ \bibnamefont
  {Cornaglia}}, \bibinfo {author} {\bibfnamefont {D.~R.}\ \bibnamefont
  {Grempel}},\ and\ \bibinfo {author} {\bibfnamefont {H.}~\bibnamefont
  {Ness}},\ }\bibfield  {title} {\bibinfo {title} {Quantum transport through a
  deformable molecular transistor},\ }\href
  {https://doi.org/10.1103/PhysRevB.71.075320} {\bibfield  {journal} {\bibinfo
  {journal} {Phys. Rev. B}\ }\textbf {\bibinfo {volume} {71}},\ \bibinfo
  {pages} {075320} (\bibinfo {year} {2005})}\BibitemShut {NoStop}%
\bibitem [{\citenamefont {Laakso}\ \emph {et~al.}(2014)\citenamefont {Laakso},
  \citenamefont {Kennes}, \citenamefont {Jakobs},\ and\ \citenamefont
  {Meden}}]{LaaksoMeden2014}%
  \BibitemOpen
  \bibfield  {author} {\bibinfo {author} {\bibfnamefont {M.~A.}\ \bibnamefont
  {Laakso}}, \bibinfo {author} {\bibfnamefont {D.~M.}\ \bibnamefont {Kennes}},
  \bibinfo {author} {\bibfnamefont {S.~G.}\ \bibnamefont {Jakobs}},\ and\
  \bibinfo {author} {\bibfnamefont {V.}~\bibnamefont {Meden}},\ }\bibfield
  {title} {\bibinfo {title} {Functional renormalization group study of the
  anderson–holstein model},\ }\href
  {https://doi.org/10.1088/1367-2630/16/2/023007} {\bibfield  {journal}
  {\bibinfo  {journal} {New Journal of Physics}\ }\textbf {\bibinfo {volume}
  {16}},\ \bibinfo {pages} {023007} (\bibinfo {year} {2014})}\BibitemShut
  {NoStop}%
\bibitem [{\citenamefont {Tanimura}\ and\ \citenamefont
  {Kubo}(1989)}]{YoshitakaKubo1989}%
  \BibitemOpen
  \bibfield  {author} {\bibinfo {author} {\bibfnamefont {Y.}~\bibnamefont
  {Tanimura}}\ and\ \bibinfo {author} {\bibfnamefont {R.}~\bibnamefont
  {Kubo}},\ }\bibfield  {title} {\bibinfo {title} {Time evolution of a quantum
  system in contact with a nearly gaussian-markoffian noise bath},\ }\href
  {https://doi.org/10.1143/JPSJ.58.101} {\bibfield  {journal} {\bibinfo
  {journal} {J. Phys. Soc. Jpn.}\ }\textbf {\bibinfo {volume} {58}},\ \bibinfo
  {pages} {101} (\bibinfo {year} {1989})}\BibitemShut {NoStop}%
\bibitem [{\citenamefont {Jin}\ \emph {et~al.}(2007)\citenamefont {Jin},
  \citenamefont {Welack}, \citenamefont {Luo}, \citenamefont {Li},
  \citenamefont {Cui}, \citenamefont {Xu},\ and\ \citenamefont
  {Yan}}]{jin2007-dynamics}%
  \BibitemOpen
  \bibfield  {author} {\bibinfo {author} {\bibfnamefont {J.}~\bibnamefont
  {Jin}}, \bibinfo {author} {\bibfnamefont {S.}~\bibnamefont {Welack}},
  \bibinfo {author} {\bibfnamefont {J.}~\bibnamefont {Luo}}, \bibinfo {author}
  {\bibfnamefont {X.-Q.}\ \bibnamefont {Li}}, \bibinfo {author} {\bibfnamefont
  {P.}~\bibnamefont {Cui}}, \bibinfo {author} {\bibfnamefont {R.-X.}\
  \bibnamefont {Xu}},\ and\ \bibinfo {author} {\bibfnamefont {Y.}~\bibnamefont
  {Yan}},\ }\bibfield  {title} {\bibinfo {title} {Dynamics of quantum
  dissipation systems interacting with fermion and boson grand canonical bath
  ensembles: Hierarchical equations of motion approach},\ }\href
  {https://doi.org/10.1063/1.2713104} {\bibfield  {journal} {\bibinfo
  {journal} {J. Chem. Phys.}\ }\textbf {\bibinfo {volume} {126}},\ \bibinfo
  {pages} {134113} (\bibinfo {year} {2007})}\BibitemShut {NoStop}%
\bibitem [{\citenamefont {Jin}\ \emph {et~al.}(2008)\citenamefont {Jin},
  \citenamefont {Zheng},\ and\ \citenamefont {Yan}}]{jin2008-exact}%
  \BibitemOpen
  \bibfield  {author} {\bibinfo {author} {\bibfnamefont {J.}~\bibnamefont
  {Jin}}, \bibinfo {author} {\bibfnamefont {X.}~\bibnamefont {Zheng}},\ and\
  \bibinfo {author} {\bibfnamefont {Y.}~\bibnamefont {Yan}},\ }\bibfield
  {title} {\bibinfo {title} {Exact dynamics of dissipative electronic systems
  and quantum transport: Hierarchical equations of motion approach},\ }\href
  {https://doi.org/10.1063/1.2938087} {\bibfield  {journal} {\bibinfo
  {journal} {J. Chem. Phys.}\ }\textbf {\bibinfo {volume} {128}},\ \bibinfo
  {pages} {234703} (\bibinfo {year} {2008})}\BibitemShut {NoStop}%
\bibitem [{\citenamefont {Yan}\ \emph {et~al.}(2016)\citenamefont {Yan},
  \citenamefont {Jin}, \citenamefont {Xu},\ and\ \citenamefont
  {Zheng}}]{yan2016-dissipation}%
  \BibitemOpen
  \bibfield  {author} {\bibinfo {author} {\bibfnamefont {Y.}~\bibnamefont
  {Yan}}, \bibinfo {author} {\bibfnamefont {J.}~\bibnamefont {Jin}}, \bibinfo
  {author} {\bibfnamefont {R.-X.}\ \bibnamefont {Xu}},\ and\ \bibinfo {author}
  {\bibfnamefont {X.}~\bibnamefont {Zheng}},\ }\bibfield  {title} {\bibinfo
  {title} {Dissipation equation of motion approach to open quantum systems},\
  }\href {https://doi.org/10.1007/s11467-016-0513-5} {\bibfield  {journal}
  {\bibinfo  {journal} {Front. Phys.}\ }\textbf {\bibinfo {volume} {11}},\
  \bibinfo {pages} {110306} (\bibinfo {year} {2016})}\BibitemShut {NoStop}%
\bibitem [{\citenamefont {Cao}\ \emph {et~al.}(2023)\citenamefont {Cao},
  \citenamefont {Ye}, \citenamefont {Xu}, \citenamefont {Zheng},\ and\
  \citenamefont {Yan}}]{cao2023-recent}%
  \BibitemOpen
  \bibfield  {author} {\bibinfo {author} {\bibfnamefont {J.}~\bibnamefont
  {Cao}}, \bibinfo {author} {\bibfnamefont {L.}~\bibnamefont {Ye}}, \bibinfo
  {author} {\bibfnamefont {R.}~\bibnamefont {Xu}}, \bibinfo {author}
  {\bibfnamefont {X.}~\bibnamefont {Zheng}},\ and\ \bibinfo {author}
  {\bibfnamefont {Y.}~\bibnamefont {Yan}},\ }\bibfield  {title} {\bibinfo
  {title} {Recent advances in fermionic hierarchical equations of motion method
  for strongly correlated quantum impurity systems},\ }\href
  {https://doi.org/10.52396/justc-2022-0164} {\bibfield  {journal} {\bibinfo
  {journal} {JUSTC}\ }\textbf {\bibinfo {volume} {53}},\ \bibinfo {pages}
  {0302} (\bibinfo {year} {2023})}\BibitemShut {NoStop}%
\bibitem [{\citenamefont {Shi}\ \emph {et~al.}(2018)\citenamefont {Shi},
  \citenamefont {Xu}, \citenamefont {Yan},\ and\ \citenamefont
  {Xu}}]{ShiXu2018}%
  \BibitemOpen
  \bibfield  {author} {\bibinfo {author} {\bibfnamefont {Q.}~\bibnamefont
  {Shi}}, \bibinfo {author} {\bibfnamefont {Y.}~\bibnamefont {Xu}}, \bibinfo
  {author} {\bibfnamefont {Y.}~\bibnamefont {Yan}},\ and\ \bibinfo {author}
  {\bibfnamefont {M.}~\bibnamefont {Xu}},\ }\bibfield  {title} {\bibinfo
  {title} {Efficient propagation of the hierarchical equations of motion using
  the matrix product state method},\ }\href {https://doi.org/10.1063/1.5026753}
  {\bibfield  {journal} {\bibinfo  {journal} {The Journal of Chemical Physics}\
  }\textbf {\bibinfo {volume} {148}},\ \bibinfo {pages} {174102} (\bibinfo
  {year} {2018})},\ \Eprint
  {https://arxiv.org/abs/https://pubs.aip.org/aip/jcp/article-pdf/doi/10.1063/1.5026753/15539991/174102\_1\_online.pdf}
  {https://pubs.aip.org/aip/jcp/article-pdf/doi/10.1063/1.5026753/15539991/174102\_1\_online.pdf}
  \BibitemShut {NoStop}%
\bibitem [{\citenamefont {Yan}\ \emph {et~al.}(2021)\citenamefont {Yan},
  \citenamefont {Xu}, \citenamefont {Li},\ and\ \citenamefont
  {Shi}}]{YanShi2021}%
  \BibitemOpen
  \bibfield  {author} {\bibinfo {author} {\bibfnamefont {Y.}~\bibnamefont
  {Yan}}, \bibinfo {author} {\bibfnamefont {M.}~\bibnamefont {Xu}}, \bibinfo
  {author} {\bibfnamefont {T.}~\bibnamefont {Li}},\ and\ \bibinfo {author}
  {\bibfnamefont {Q.}~\bibnamefont {Shi}},\ }\bibfield  {title} {\bibinfo
  {title} {Efficient propagation of the hierarchical equations of motion using
  the tucker and hierarchical tucker tensors},\ }\href
  {https://doi.org/10.1063/5.0050720} {\bibfield  {journal} {\bibinfo
  {journal} {The Journal of Chemical Physics}\ }\textbf {\bibinfo {volume}
  {154}},\ \bibinfo {pages} {194104} (\bibinfo {year} {2021})},\ \Eprint
  {https://arxiv.org/abs/https://pubs.aip.org/aip/jcp/article-pdf/doi/10.1063/5.0050720/15977717/194104\_1\_online.pdf}
  {https://pubs.aip.org/aip/jcp/article-pdf/doi/10.1063/5.0050720/15977717/194104\_1\_online.pdf}
  \BibitemShut {NoStop}%
\bibitem [{\citenamefont {Dan}\ \emph {et~al.}(2023)\citenamefont {Dan},
  \citenamefont {Xu}, \citenamefont {Stockburger}, \citenamefont {Ankerhold},\
  and\ \citenamefont {Shi}}]{DanShi2023}%
  \BibitemOpen
  \bibfield  {author} {\bibinfo {author} {\bibfnamefont {X.}~\bibnamefont
  {Dan}}, \bibinfo {author} {\bibfnamefont {M.}~\bibnamefont {Xu}}, \bibinfo
  {author} {\bibfnamefont {J.~T.}\ \bibnamefont {Stockburger}}, \bibinfo
  {author} {\bibfnamefont {J.}~\bibnamefont {Ankerhold}},\ and\ \bibinfo
  {author} {\bibfnamefont {Q.}~\bibnamefont {Shi}},\ }\bibfield  {title}
  {\bibinfo {title} {Efficient low-temperature simulations for fermionic
  reservoirs with the hierarchical equations of motion method: Application to
  the anderson impurity model},\ }\href
  {https://doi.org/10.1103/PhysRevB.107.195429} {\bibfield  {journal} {\bibinfo
   {journal} {Phys. Rev. B}\ }\textbf {\bibinfo {volume} {107}},\ \bibinfo
  {pages} {195429} (\bibinfo {year} {2023})}\BibitemShut {NoStop}%
\bibitem [{\citenamefont {Wolf}\ \emph
  {et~al.}(2014{\natexlab{a}})\citenamefont {Wolf}, \citenamefont {McCulloch},
  \citenamefont {Parcollet},\ and\ \citenamefont
  {Schollw\"ock}}]{WolfSchollwock2014b}%
  \BibitemOpen
  \bibfield  {author} {\bibinfo {author} {\bibfnamefont {F.~A.}\ \bibnamefont
  {Wolf}}, \bibinfo {author} {\bibfnamefont {I.~P.}\ \bibnamefont {McCulloch}},
  \bibinfo {author} {\bibfnamefont {O.}~\bibnamefont {Parcollet}},\ and\
  \bibinfo {author} {\bibfnamefont {U.}~\bibnamefont {Schollw\"ock}},\
  }\bibfield  {title} {\bibinfo {title} {Chebyshev matrix product state
  impurity solver for dynamical mean-field theory},\ }\href
  {https://doi.org/10.1103/PhysRevB.90.115124} {\bibfield  {journal} {\bibinfo
  {journal} {Phys. Rev. B}\ }\textbf {\bibinfo {volume} {90}},\ \bibinfo
  {pages} {115124} (\bibinfo {year} {2014}{\natexlab{a}})}\BibitemShut
  {NoStop}%
\bibitem [{\citenamefont {Ganahl}\ \emph {et~al.}(2014)\citenamefont {Ganahl},
  \citenamefont {Thunstr\"om}, \citenamefont {Verstraete}, \citenamefont
  {Held},\ and\ \citenamefont {Evertz}}]{GanahlEvertz2014}%
  \BibitemOpen
  \bibfield  {author} {\bibinfo {author} {\bibfnamefont {M.}~\bibnamefont
  {Ganahl}}, \bibinfo {author} {\bibfnamefont {P.}~\bibnamefont {Thunstr\"om}},
  \bibinfo {author} {\bibfnamefont {F.}~\bibnamefont {Verstraete}}, \bibinfo
  {author} {\bibfnamefont {K.}~\bibnamefont {Held}},\ and\ \bibinfo {author}
  {\bibfnamefont {H.~G.}\ \bibnamefont {Evertz}},\ }\bibfield  {title}
  {\bibinfo {title} {Chebyshev expansion for impurity models using matrix
  product states},\ }\href {https://doi.org/10.1103/PhysRevB.90.045144}
  {\bibfield  {journal} {\bibinfo  {journal} {Phys. Rev. B}\ }\textbf {\bibinfo
  {volume} {90}},\ \bibinfo {pages} {045144} (\bibinfo {year}
  {2014})}\BibitemShut {NoStop}%
\bibitem [{\citenamefont {Ganahl}\ \emph {et~al.}(2015)\citenamefont {Ganahl},
  \citenamefont {Aichhorn}, \citenamefont {Evertz}, \citenamefont
  {Thunstr\"om}, \citenamefont {Held},\ and\ \citenamefont
  {Verstraete}}]{GanahlVerstraete2015}%
  \BibitemOpen
  \bibfield  {author} {\bibinfo {author} {\bibfnamefont {M.}~\bibnamefont
  {Ganahl}}, \bibinfo {author} {\bibfnamefont {M.}~\bibnamefont {Aichhorn}},
  \bibinfo {author} {\bibfnamefont {H.~G.}\ \bibnamefont {Evertz}}, \bibinfo
  {author} {\bibfnamefont {P.}~\bibnamefont {Thunstr\"om}}, \bibinfo {author}
  {\bibfnamefont {K.}~\bibnamefont {Held}},\ and\ \bibinfo {author}
  {\bibfnamefont {F.}~\bibnamefont {Verstraete}},\ }\bibfield  {title}
  {\bibinfo {title} {Efficient dmft impurity solver using real-time dynamics
  with matrix product states},\ }\href
  {https://doi.org/10.1103/PhysRevB.92.155132} {\bibfield  {journal} {\bibinfo
  {journal} {Phys. Rev. B}\ }\textbf {\bibinfo {volume} {92}},\ \bibinfo
  {pages} {155132} (\bibinfo {year} {2015})}\BibitemShut {NoStop}%
\bibitem [{\citenamefont {Wolf}\ \emph {et~al.}(2015)\citenamefont {Wolf},
  \citenamefont {Go}, \citenamefont {McCulloch}, \citenamefont {Millis},\ and\
  \citenamefont {Schollw\"ock}}]{WolfSchollwock2015}%
  \BibitemOpen
  \bibfield  {author} {\bibinfo {author} {\bibfnamefont {F.~A.}\ \bibnamefont
  {Wolf}}, \bibinfo {author} {\bibfnamefont {A.}~\bibnamefont {Go}}, \bibinfo
  {author} {\bibfnamefont {I.~P.}\ \bibnamefont {McCulloch}}, \bibinfo {author}
  {\bibfnamefont {A.~J.}\ \bibnamefont {Millis}},\ and\ \bibinfo {author}
  {\bibfnamefont {U.}~\bibnamefont {Schollw\"ock}},\ }\bibfield  {title}
  {\bibinfo {title} {Imaginary-time matrix product state impurity solver for
  dynamical mean-field theory},\ }\href
  {https://doi.org/10.1103/PhysRevX.5.041032} {\bibfield  {journal} {\bibinfo
  {journal} {Phys. Rev. X}\ }\textbf {\bibinfo {volume} {5}},\ \bibinfo {pages}
  {041032} (\bibinfo {year} {2015})}\BibitemShut {NoStop}%
\bibitem [{\citenamefont {Garc\'{\i}a}\ \emph {et~al.}(2004)\citenamefont
  {Garc\'{\i}a}, \citenamefont {Hallberg},\ and\ \citenamefont
  {Rozenberg}}]{GarciaRozenberg2004}%
  \BibitemOpen
  \bibfield  {author} {\bibinfo {author} {\bibfnamefont {D.~J.}\ \bibnamefont
  {Garc\'{\i}a}}, \bibinfo {author} {\bibfnamefont {K.}~\bibnamefont
  {Hallberg}},\ and\ \bibinfo {author} {\bibfnamefont {M.~J.}\ \bibnamefont
  {Rozenberg}},\ }\bibfield  {title} {\bibinfo {title} {Dynamical mean field
  theory with the density matrix renormalization group},\ }\href
  {https://doi.org/10.1103/PhysRevLett.93.246403} {\bibfield  {journal}
  {\bibinfo  {journal} {Phys. Rev. Lett.}\ }\textbf {\bibinfo {volume} {93}},\
  \bibinfo {pages} {246403} (\bibinfo {year} {2004})}\BibitemShut {NoStop}%
\bibitem [{\citenamefont {Nishimoto}\ \emph {et~al.}(2006)\citenamefont
  {Nishimoto}, \citenamefont {Gebhard},\ and\ \citenamefont
  {Jeckelmann}}]{NishimotoJeckelmann2006}%
  \BibitemOpen
  \bibfield  {author} {\bibinfo {author} {\bibfnamefont {S.}~\bibnamefont
  {Nishimoto}}, \bibinfo {author} {\bibfnamefont {F.}~\bibnamefont {Gebhard}},\
  and\ \bibinfo {author} {\bibfnamefont {E.}~\bibnamefont {Jeckelmann}},\
  }\bibfield  {title} {\bibinfo {title} {Dynamical mean-field theory
  calculation with the dynamical density-matrix renormalization group},\ }\href
  {https://doi.org/https://doi.org/10.1016/j.physb.2006.01.104} {\bibfield
  {journal} {\bibinfo  {journal} {Physica B Condens. Matter}\ }\textbf
  {\bibinfo {volume} {378-380}},\ \bibinfo {pages} {283} (\bibinfo {year}
  {2006})}\BibitemShut {NoStop}%
\bibitem [{\citenamefont {Weichselbaum}\ \emph {et~al.}(2009)\citenamefont
  {Weichselbaum}, \citenamefont {Verstraete}, \citenamefont {Schollw\"ock},
  \citenamefont {Cirac},\ and\ \citenamefont {von
  Delft}}]{WeichselbaumDelft2009}%
  \BibitemOpen
  \bibfield  {author} {\bibinfo {author} {\bibfnamefont {A.}~\bibnamefont
  {Weichselbaum}}, \bibinfo {author} {\bibfnamefont {F.}~\bibnamefont
  {Verstraete}}, \bibinfo {author} {\bibfnamefont {U.}~\bibnamefont
  {Schollw\"ock}}, \bibinfo {author} {\bibfnamefont {J.~I.}\ \bibnamefont
  {Cirac}},\ and\ \bibinfo {author} {\bibfnamefont {J.}~\bibnamefont {von
  Delft}},\ }\bibfield  {title} {\bibinfo {title} {Variational
  matrix-product-state approach to quantum impurity models},\ }\href
  {https://doi.org/10.1103/PhysRevB.80.165117} {\bibfield  {journal} {\bibinfo
  {journal} {Phys. Rev. B}\ }\textbf {\bibinfo {volume} {80}},\ \bibinfo
  {pages} {165117} (\bibinfo {year} {2009})}\BibitemShut {NoStop}%
\bibitem [{\citenamefont {Bauernfeind}\ \emph {et~al.}(2017)\citenamefont
  {Bauernfeind}, \citenamefont {Zingl}, \citenamefont {Triebl}, \citenamefont
  {Aichhorn},\ and\ \citenamefont {Evertz}}]{BauernfeindEvertz2017}%
  \BibitemOpen
  \bibfield  {author} {\bibinfo {author} {\bibfnamefont {D.}~\bibnamefont
  {Bauernfeind}}, \bibinfo {author} {\bibfnamefont {M.}~\bibnamefont {Zingl}},
  \bibinfo {author} {\bibfnamefont {R.}~\bibnamefont {Triebl}}, \bibinfo
  {author} {\bibfnamefont {M.}~\bibnamefont {Aichhorn}},\ and\ \bibinfo
  {author} {\bibfnamefont {H.~G.}\ \bibnamefont {Evertz}},\ }\bibfield  {title}
  {\bibinfo {title} {Fork tensor-product states: Efficient multiorbital
  real-time dmft solver},\ }\href {https://doi.org/10.1103/PhysRevX.7.031013}
  {\bibfield  {journal} {\bibinfo  {journal} {Phys. Rev. X}\ }\textbf {\bibinfo
  {volume} {7}},\ \bibinfo {pages} {031013} (\bibinfo {year}
  {2017})}\BibitemShut {NoStop}%
\bibitem [{\citenamefont {Werner}\ \emph {et~al.}(2023)\citenamefont {Werner},
  \citenamefont {Lotze},\ and\ \citenamefont {Arrigoni}}]{WernerArrigoni2023}%
  \BibitemOpen
  \bibfield  {author} {\bibinfo {author} {\bibfnamefont {D.}~\bibnamefont
  {Werner}}, \bibinfo {author} {\bibfnamefont {J.}~\bibnamefont {Lotze}},\ and\
  \bibinfo {author} {\bibfnamefont {E.}~\bibnamefont {Arrigoni}},\ }\bibfield
  {title} {\bibinfo {title} {Configuration interaction based nonequilibrium
  steady state impurity solver},\ }\href
  {https://doi.org/10.1103/PhysRevB.107.075119} {\bibfield  {journal} {\bibinfo
   {journal} {Phys. Rev. B}\ }\textbf {\bibinfo {volume} {107}},\ \bibinfo
  {pages} {075119} (\bibinfo {year} {2023})}\BibitemShut {NoStop}%
\bibitem [{\citenamefont {Kohn}\ and\ \citenamefont
  {Santoro}(2021)}]{KohnSantoro2021}%
  \BibitemOpen
  \bibfield  {author} {\bibinfo {author} {\bibfnamefont {L.}~\bibnamefont
  {Kohn}}\ and\ \bibinfo {author} {\bibfnamefont {G.~E.}\ \bibnamefont
  {Santoro}},\ }\bibfield  {title} {\bibinfo {title} {Efficient mapping for
  anderson impurity problems with matrix product states},\ }\href
  {https://doi.org/10.1103/PhysRevB.104.014303} {\bibfield  {journal} {\bibinfo
   {journal} {Phys. Rev. B}\ }\textbf {\bibinfo {volume} {104}},\ \bibinfo
  {pages} {014303} (\bibinfo {year} {2021})}\BibitemShut {NoStop}%
\bibitem [{\citenamefont {Kohn}\ and\ \citenamefont
  {Santoro}(2022)}]{KohnSantoro2022}%
  \BibitemOpen
  \bibfield  {author} {\bibinfo {author} {\bibfnamefont {L.}~\bibnamefont
  {Kohn}}\ and\ \bibinfo {author} {\bibfnamefont {G.~E.}\ \bibnamefont
  {Santoro}},\ }\bibfield  {title} {\bibinfo {title} {Quench dynamics of the
  anderson impurity model at finite temperature using matrix product states:
  entanglement and bath dynamics},\ }\href
  {https://doi.org/10.1088/1742-5468/ac729b} {\bibfield  {journal} {\bibinfo
  {journal} {J. Stat. Mech. Theory Exp.}\ }\textbf {\bibinfo {volume} {2022}},\
  \bibinfo {pages} {063102} (\bibinfo {year} {2022})}\BibitemShut {NoStop}%
\bibitem [{\citenamefont {Xu}\ \emph {et~al.}(2024)\citenamefont {Xu},
  \citenamefont {Guo},\ and\ \citenamefont {Chen}}]{XuChen2024}%
  \BibitemOpen
  \bibfield  {author} {\bibinfo {author} {\bibfnamefont {X.}~\bibnamefont
  {Xu}}, \bibinfo {author} {\bibfnamefont {C.}~\bibnamefont {Guo}},\ and\
  \bibinfo {author} {\bibfnamefont {R.}~\bibnamefont {Chen}},\ }\bibfield
  {title} {\bibinfo {title} {Grassmann time-evolving matrix product operators:
  An efficient numerical approach for fermionic path integral simulations},\
  }\href {https://doi.org/10.1063/5.0226167} {\bibfield  {journal} {\bibinfo
  {journal} {J. Chem. Phys.}\ }\textbf {\bibinfo {volume} {161}},\ \bibinfo
  {pages} {151001} (\bibinfo {year} {2024})}\BibitemShut {NoStop}%
\bibitem [{\citenamefont {Strathearn}\ \emph {et~al.}(2018)\citenamefont
  {Strathearn}, \citenamefont {Kirton}, \citenamefont {Kilda}, \citenamefont
  {Keeling},\ and\ \citenamefont {Lovett}}]{StrathearnLovett2018}%
  \BibitemOpen
  \bibfield  {author} {\bibinfo {author} {\bibfnamefont {A.}~\bibnamefont
  {Strathearn}}, \bibinfo {author} {\bibfnamefont {P.}~\bibnamefont {Kirton}},
  \bibinfo {author} {\bibfnamefont {D.}~\bibnamefont {Kilda}}, \bibinfo
  {author} {\bibfnamefont {J.}~\bibnamefont {Keeling}},\ and\ \bibinfo {author}
  {\bibfnamefont {B.~W.}\ \bibnamefont {Lovett}},\ }\bibfield  {title}
  {\bibinfo {title} {Efficient non-markovian quantum dynamics using
  time-evolving matrix product operators},\ }\href
  {https://doi.org/10.1038/s41467-018-05617-3} {\bibfield  {journal} {\bibinfo
  {journal} {Nat. Commun.}\ }\textbf {\bibinfo {volume} {9}},\ \bibinfo {pages}
  {3322} (\bibinfo {year} {2018})}\BibitemShut {NoStop}%
\bibitem [{\citenamefont {Feynman}\ and\ \citenamefont
  {Vernon}(1963)}]{FeynmanVernon1963}%
  \BibitemOpen
  \bibfield  {author} {\bibinfo {author} {\bibfnamefont {R.~P.}\ \bibnamefont
  {Feynman}}\ and\ \bibinfo {author} {\bibfnamefont {F.~L.}\ \bibnamefont
  {Vernon}},\ }\bibfield  {title} {\bibinfo {title} {The theory of a general
  quantum system interacting with a linear dissipative system},\ }\href
  {https://doi.org/10.1016/0003-4916(63)90068-X} {\bibfield  {journal}
  {\bibinfo  {journal} {Ann. Phys.}\ }\textbf {\bibinfo {volume} {24}},\
  \bibinfo {pages} {118} (\bibinfo {year} {1963})}\BibitemShut {NoStop}%
\bibitem [{\citenamefont {Chen}\ \emph
  {et~al.}(2024{\natexlab{a}})\citenamefont {Chen}, \citenamefont {Xu},\ and\
  \citenamefont {Guo}}]{ChenGuo2024b}%
  \BibitemOpen
  \bibfield  {author} {\bibinfo {author} {\bibfnamefont {R.}~\bibnamefont
  {Chen}}, \bibinfo {author} {\bibfnamefont {X.}~\bibnamefont {Xu}},\ and\
  \bibinfo {author} {\bibfnamefont {C.}~\bibnamefont {Guo}},\ }\bibfield
  {title} {\bibinfo {title} {Grassmann time-evolving matrix product operators
  for equilibrium quantum impurity problems},\ }\href
  {https://doi.org/10.1088/1367-2630/ad19fa} {\bibfield  {journal} {\bibinfo
  {journal} {New J. Phys.}\ }\textbf {\bibinfo {volume} {26}},\ \bibinfo
  {pages} {013019} (\bibinfo {year} {2024}{\natexlab{a}})}\BibitemShut
  {NoStop}%
\bibitem [{\citenamefont {Chen}\ \emph
  {et~al.}(2024{\natexlab{b}})\citenamefont {Chen}, \citenamefont {Xu},\ and\
  \citenamefont {Guo}}]{ChenGuo2024a}%
  \BibitemOpen
  \bibfield  {author} {\bibinfo {author} {\bibfnamefont {R.}~\bibnamefont
  {Chen}}, \bibinfo {author} {\bibfnamefont {X.}~\bibnamefont {Xu}},\ and\
  \bibinfo {author} {\bibfnamefont {C.}~\bibnamefont {Guo}},\ }\bibfield
  {title} {\bibinfo {title} {Grassmann time-evolving matrix product operators
  for quantum impurity models},\ }\href
  {https://doi.org/10.1103/PhysRevB.109.045140} {\bibfield  {journal} {\bibinfo
   {journal} {Phys. Rev. B}\ }\textbf {\bibinfo {volume} {109}},\ \bibinfo
  {pages} {045140} (\bibinfo {year} {2024}{\natexlab{b}})}\BibitemShut
  {NoStop}%
\bibitem [{\citenamefont {Chen}\ \emph
  {et~al.}(2024{\natexlab{c}})\citenamefont {Chen}, \citenamefont {Xu},\ and\
  \citenamefont {Guo}}]{ChenGuo2024c}%
  \BibitemOpen
  \bibfield  {author} {\bibinfo {author} {\bibfnamefont {R.}~\bibnamefont
  {Chen}}, \bibinfo {author} {\bibfnamefont {X.}~\bibnamefont {Xu}},\ and\
  \bibinfo {author} {\bibfnamefont {C.}~\bibnamefont {Guo}},\ }\bibfield
  {title} {\bibinfo {title} {Real-time impurity solver using grassmann
  time-evolving matrix product operators},\ }\href
  {https://doi.org/10.1103/PhysRevB.109.165113} {\bibfield  {journal} {\bibinfo
   {journal} {Phys. Rev. B}\ }\textbf {\bibinfo {volume} {109}},\ \bibinfo
  {pages} {165113} (\bibinfo {year} {2024}{\natexlab{c}})}\BibitemShut
  {NoStop}%
\bibitem [{\citenamefont {Chen}\ and\ \citenamefont
  {Guo}(2024)}]{ChenGuo2024g}%
  \BibitemOpen
  \bibfield  {author} {\bibinfo {author} {\bibfnamefont {R.}~\bibnamefont
  {Chen}}\ and\ \bibinfo {author} {\bibfnamefont {C.}~\bibnamefont {Guo}},\
  }\bibfield  {title} {\bibinfo {title} {Solving equilibrium quantum impurity
  problems on the l-shaped kadanoff-baym contour},\ }\href
  {https://doi.org/10.1103/PhysRevB.110.165114} {\bibfield  {journal} {\bibinfo
   {journal} {Phys. Rev. B}\ }\textbf {\bibinfo {volume} {110}},\ \bibinfo
  {pages} {165114} (\bibinfo {year} {2024})}\BibitemShut {NoStop}%
\bibitem [{\citenamefont {Sun}\ \emph {et~al.}(2025)\citenamefont {Sun},
  \citenamefont {Chen}, \citenamefont {Li},\ and\ \citenamefont
  {Guo}}]{SunGuo2025}%
  \BibitemOpen
  \bibfield  {author} {\bibinfo {author} {\bibfnamefont {Z.}~\bibnamefont
  {Sun}}, \bibinfo {author} {\bibfnamefont {R.}~\bibnamefont {Chen}}, \bibinfo
  {author} {\bibfnamefont {Z.}~\bibnamefont {Li}},\ and\ \bibinfo {author}
  {\bibfnamefont {C.}~\bibnamefont {Guo}},\ }\bibfield  {title} {\bibinfo
  {title} {Infinite grassmann time-evolving matrix product operators for
  quantum impurity problems after a quench},\ }\href
  {https://doi.org/10.1103/g5y8-jfb2} {\bibfield  {journal} {\bibinfo
  {journal} {Phys. Rev. B}\ }\textbf {\bibinfo {volume} {112}},\ \bibinfo
  {pages} {125145} (\bibinfo {year} {2025})}\BibitemShut {NoStop}%
\bibitem [{\citenamefont {Orth}\ \emph {et~al.}(2013)\citenamefont {Orth},
  \citenamefont {Imambekov},\ and\ \citenamefont {Le~Hur}}]{OrthLe2013}%
  \BibitemOpen
  \bibfield  {author} {\bibinfo {author} {\bibfnamefont {P.~P.}\ \bibnamefont
  {Orth}}, \bibinfo {author} {\bibfnamefont {A.}~\bibnamefont {Imambekov}},\
  and\ \bibinfo {author} {\bibfnamefont {K.}~\bibnamefont {Le~Hur}},\
  }\bibfield  {title} {\bibinfo {title} {Nonperturbative stochastic method for
  driven spin-boson model},\ }\href
  {https://doi.org/10.1103/PhysRevB.87.014305} {\bibfield  {journal} {\bibinfo
  {journal} {Phys. Rev. B}\ }\textbf {\bibinfo {volume} {87}},\ \bibinfo
  {pages} {014305} (\bibinfo {year} {2013})}\BibitemShut {NoStop}%
\bibitem [{\citenamefont {Kamar}\ \emph {et~al.}(2024)\citenamefont {Kamar},
  \citenamefont {Paz},\ and\ \citenamefont {Maghrebi}}]{KamarMaghrebi2024}%
  \BibitemOpen
  \bibfield  {author} {\bibinfo {author} {\bibfnamefont {N.~A.}\ \bibnamefont
  {Kamar}}, \bibinfo {author} {\bibfnamefont {D.~A.}\ \bibnamefont {Paz}},\
  and\ \bibinfo {author} {\bibfnamefont {M.~F.}\ \bibnamefont {Maghrebi}},\
  }\bibfield  {title} {\bibinfo {title} {Spin-boson model under dephasing:
  Markovian versus non-markovian dynamics},\ }\href
  {https://doi.org/10.1103/PhysRevB.110.075126} {\bibfield  {journal} {\bibinfo
   {journal} {Phys. Rev. B}\ }\textbf {\bibinfo {volume} {110}},\ \bibinfo
  {pages} {075126} (\bibinfo {year} {2024})}\BibitemShut {NoStop}%
\bibitem [{\citenamefont {Schollwöck}(2011)}]{Schollwock2011}%
  \BibitemOpen
  \bibfield  {author} {\bibinfo {author} {\bibfnamefont {U.}~\bibnamefont
  {Schollwöck}},\ }\bibfield  {title} {\bibinfo {title} {The density-matrix
  renormalization group in the age of matrix product states},\ }\href
  {https://doi.org/https://doi.org/10.1016/j.aop.2010.09.012} {\bibfield
  {journal} {\bibinfo  {journal} {Ann. Phys.}\ }\textbf {\bibinfo {volume}
  {326}},\ \bibinfo {pages} {96} (\bibinfo {year} {2011})}\BibitemShut
  {NoStop}%
\bibitem [{\citenamefont {Fux}\ \emph {et~al.}(2023)\citenamefont {Fux},
  \citenamefont {Kilda}, \citenamefont {Lovett},\ and\ \citenamefont
  {Keeling}}]{FuxKeeling2023}%
  \BibitemOpen
  \bibfield  {author} {\bibinfo {author} {\bibfnamefont {G.~E.}\ \bibnamefont
  {Fux}}, \bibinfo {author} {\bibfnamefont {D.}~\bibnamefont {Kilda}}, \bibinfo
  {author} {\bibfnamefont {B.~W.}\ \bibnamefont {Lovett}},\ and\ \bibinfo
  {author} {\bibfnamefont {J.}~\bibnamefont {Keeling}},\ }\bibfield  {title}
  {\bibinfo {title} {Tensor network simulation of chains of non-markovian open
  quantum systems},\ }\href {https://doi.org/10.1103/PhysRevResearch.5.033078}
  {\bibfield  {journal} {\bibinfo  {journal} {Phys. Rev. Res.}\ }\textbf
  {\bibinfo {volume} {5}},\ \bibinfo {pages} {033078} (\bibinfo {year}
  {2023})}\BibitemShut {NoStop}%
\bibitem [{\citenamefont {Wolf}\ \emph
  {et~al.}(2014{\natexlab{b}})\citenamefont {Wolf}, \citenamefont {McCulloch},\
  and\ \citenamefont {Schollw\"ock}}]{WolfSchollwock2014}%
  \BibitemOpen
  \bibfield  {author} {\bibinfo {author} {\bibfnamefont {F.~A.}\ \bibnamefont
  {Wolf}}, \bibinfo {author} {\bibfnamefont {I.~P.}\ \bibnamefont
  {McCulloch}},\ and\ \bibinfo {author} {\bibfnamefont {U.}~\bibnamefont
  {Schollw\"ock}},\ }\bibfield  {title} {\bibinfo {title} {Solving
  nonequilibrium dynamical mean-field theory using matrix product states},\
  }\href {https://doi.org/10.1103/PhysRevB.90.235131} {\bibfield  {journal}
  {\bibinfo  {journal} {Phys. Rev. B}\ }\textbf {\bibinfo {volume} {90}},\
  \bibinfo {pages} {235131} (\bibinfo {year} {2014}{\natexlab{b}})}\BibitemShut
  {NoStop}%
\bibitem [{\citenamefont {Chen}(2025)}]{Chen2025}%
  \BibitemOpen
  \bibfield  {author} {\bibinfo {author} {\bibfnamefont {R.}~\bibnamefont
  {Chen}},\ }\bibfield  {title} {\bibinfo {title} {Path integral formalism for
  quantum open systems},\ }\href
  {https://doi.org/https://doi.org/10.1016/j.aop.2025.170083} {\bibfield
  {journal} {\bibinfo  {journal} {Ann. Phys.}\ }\textbf {\bibinfo {volume}
  {480}},\ \bibinfo {pages} {170083} (\bibinfo {year} {2025})}\BibitemShut
  {NoStop}%
\bibitem [{\citenamefont {J\o{}rgensen}\ and\ \citenamefont
  {Pollock}(2019)}]{JorgensenPollock2019}%
  \BibitemOpen
  \bibfield  {author} {\bibinfo {author} {\bibfnamefont {M.~R.}\ \bibnamefont
  {J\o{}rgensen}}\ and\ \bibinfo {author} {\bibfnamefont {F.~A.}\ \bibnamefont
  {Pollock}},\ }\bibfield  {title} {\bibinfo {title} {Exploiting the causal
  tensor network structure of quantum processes to efficiently simulate
  non-markovian path integrals},\ }\href
  {https://doi.org/10.1103/PhysRevLett.123.240602} {\bibfield  {journal}
  {\bibinfo  {journal} {Phys. Rev. Lett.}\ }\textbf {\bibinfo {volume} {123}},\
  \bibinfo {pages} {240602} (\bibinfo {year} {2019})}\BibitemShut {NoStop}%
\bibitem [{\citenamefont {Guo}\ \emph {et~al.}(2020)\citenamefont {Guo},
  \citenamefont {Modi},\ and\ \citenamefont {Poletti}}]{GuoPoletti2020}%
  \BibitemOpen
  \bibfield  {author} {\bibinfo {author} {\bibfnamefont {C.}~\bibnamefont
  {Guo}}, \bibinfo {author} {\bibfnamefont {K.}~\bibnamefont {Modi}},\ and\
  \bibinfo {author} {\bibfnamefont {D.}~\bibnamefont {Poletti}},\ }\bibfield
  {title} {\bibinfo {title} {Tensor-network-based machine learning of
  non-markovian quantum processes},\ }\href
  {https://doi.org/10.1103/PhysRevA.102.062414} {\bibfield  {journal} {\bibinfo
   {journal} {Phys. Rev. A}\ }\textbf {\bibinfo {volume} {102}},\ \bibinfo
  {pages} {062414} (\bibinfo {year} {2020})}\BibitemShut {NoStop}%
\bibitem [{\citenamefont {Zhang}\ \emph {et~al.}(2025)\citenamefont {Zhang},
  \citenamefont {Wu}, \citenamefont {White}, \citenamefont {Xiang},
  \citenamefont {Hu}, \citenamefont {Peng}, \citenamefont {Liu}, \citenamefont
  {Zheng}, \citenamefont {Fu}, \citenamefont {Huang}, \citenamefont {Poletti},
  \citenamefont {Modi}, \citenamefont {Wu}, \citenamefont {Deng},\ and\
  \citenamefont {Guo}}]{ZhangGuo2025}%
  \BibitemOpen
  \bibfield  {author} {\bibinfo {author} {\bibfnamefont {X.}~\bibnamefont
  {Zhang}}, \bibinfo {author} {\bibfnamefont {Z.}~\bibnamefont {Wu}}, \bibinfo
  {author} {\bibfnamefont {G.~A.~L.}\ \bibnamefont {White}}, \bibinfo {author}
  {\bibfnamefont {Z.}~\bibnamefont {Xiang}}, \bibinfo {author} {\bibfnamefont
  {S.}~\bibnamefont {Hu}}, \bibinfo {author} {\bibfnamefont {Z.}~\bibnamefont
  {Peng}}, \bibinfo {author} {\bibfnamefont {Y.}~\bibnamefont {Liu}}, \bibinfo
  {author} {\bibfnamefont {D.}~\bibnamefont {Zheng}}, \bibinfo {author}
  {\bibfnamefont {X.}~\bibnamefont {Fu}}, \bibinfo {author} {\bibfnamefont
  {A.}~\bibnamefont {Huang}}, \bibinfo {author} {\bibfnamefont
  {D.}~\bibnamefont {Poletti}}, \bibinfo {author} {\bibfnamefont
  {K.}~\bibnamefont {Modi}}, \bibinfo {author} {\bibfnamefont {J.}~\bibnamefont
  {Wu}}, \bibinfo {author} {\bibfnamefont {M.}~\bibnamefont {Deng}},\ and\
  \bibinfo {author} {\bibfnamefont {C.}~\bibnamefont {Guo}},\ }\bibfield
  {title} {\bibinfo {title} {Learning and forecasting open quantum dynamics
  with correlated noise},\ }\href {https://doi.org/10.1038/s42005-025-01944-2}
  {\bibfield  {journal} {\bibinfo  {journal} {Communications Physics}\ }\textbf
  {\bibinfo {volume} {8}},\ \bibinfo {pages} {29} (\bibinfo {year}
  {2025})}\BibitemShut {NoStop}%
\bibitem [{\citenamefont {Taranto}\ \emph {et~al.}(2025)\citenamefont
  {Taranto}, \citenamefont {Milz}, \citenamefont {Murao}, \citenamefont
  {Quintino},\ and\ \citenamefont {Modi}}]{TarantoModi2025}%
  \BibitemOpen
  \bibfield  {author} {\bibinfo {author} {\bibfnamefont {P.}~\bibnamefont
  {Taranto}}, \bibinfo {author} {\bibfnamefont {S.}~\bibnamefont {Milz}},
  \bibinfo {author} {\bibfnamefont {M.}~\bibnamefont {Murao}}, \bibinfo
  {author} {\bibfnamefont {M.~T.}\ \bibnamefont {Quintino}},\ and\ \bibinfo
  {author} {\bibfnamefont {K.}~\bibnamefont {Modi}},\ }\href
  {https://arxiv.org/abs/2503.09693} {\bibinfo {title} {Higher-order quantum
  operations}} (\bibinfo {year} {2025}),\ \Eprint
  {https://arxiv.org/abs/2503.09693} {arXiv:2503.09693 [quant-ph]} \BibitemShut
  {NoStop}%
\bibitem [{\citenamefont {Guo}\ and\ \citenamefont
  {Chen}(2024)}]{GuoChen2024d}%
  \BibitemOpen
  \bibfield  {author} {\bibinfo {author} {\bibfnamefont {C.}~\bibnamefont
  {Guo}}\ and\ \bibinfo {author} {\bibfnamefont {R.}~\bibnamefont {Chen}},\
  }\bibfield  {title} {\bibinfo {title} {{Efficient construction of the
  Feynman-Vernon influence functional as matrix product states}},\ }\href
  {https://doi.org/10.21468/SciPostPhysCore.7.3.063} {\bibfield  {journal}
  {\bibinfo  {journal} {SciPost Phys. Core}\ }\textbf {\bibinfo {volume} {7}},\
  \bibinfo {pages} {063} (\bibinfo {year} {2024})}\BibitemShut {NoStop}%
\bibitem [{\citenamefont {Guo}\ \emph {et~al.}(2019)\citenamefont {Guo},
  \citenamefont {Liu}, \citenamefont {Xiong}, \citenamefont {Xue},
  \citenamefont {Fu}, \citenamefont {Huang}, \citenamefont {Qiang},
  \citenamefont {Xu}, \citenamefont {Liu}, \citenamefont {Zheng}, \citenamefont
  {Huang}, \citenamefont {Deng}, \citenamefont {Poletti}, \citenamefont {Bao},\
  and\ \citenamefont {Wu}}]{GuoWu2019}%
  \BibitemOpen
  \bibfield  {author} {\bibinfo {author} {\bibfnamefont {C.}~\bibnamefont
  {Guo}}, \bibinfo {author} {\bibfnamefont {Y.}~\bibnamefont {Liu}}, \bibinfo
  {author} {\bibfnamefont {M.}~\bibnamefont {Xiong}}, \bibinfo {author}
  {\bibfnamefont {S.}~\bibnamefont {Xue}}, \bibinfo {author} {\bibfnamefont
  {X.}~\bibnamefont {Fu}}, \bibinfo {author} {\bibfnamefont {A.}~\bibnamefont
  {Huang}}, \bibinfo {author} {\bibfnamefont {X.}~\bibnamefont {Qiang}},
  \bibinfo {author} {\bibfnamefont {P.}~\bibnamefont {Xu}}, \bibinfo {author}
  {\bibfnamefont {J.}~\bibnamefont {Liu}}, \bibinfo {author} {\bibfnamefont
  {S.}~\bibnamefont {Zheng}}, \bibinfo {author} {\bibfnamefont {H.-L.}\
  \bibnamefont {Huang}}, \bibinfo {author} {\bibfnamefont {M.}~\bibnamefont
  {Deng}}, \bibinfo {author} {\bibfnamefont {D.}~\bibnamefont {Poletti}},
  \bibinfo {author} {\bibfnamefont {W.-S.}\ \bibnamefont {Bao}},\ and\ \bibinfo
  {author} {\bibfnamefont {J.}~\bibnamefont {Wu}},\ }\bibfield  {title}
  {\bibinfo {title} {General-purpose quantum circuit simulator with projected
  entangled-pair states and the quantum supremacy frontier},\ }\href
  {https://doi.org/10.1103/PhysRevLett.123.190501} {\bibfield  {journal}
  {\bibinfo  {journal} {Phys. Rev. Lett.}\ }\textbf {\bibinfo {volume} {123}},\
  \bibinfo {pages} {190501} (\bibinfo {year} {2019})}\BibitemShut {NoStop}%
\bibitem [{\citenamefont {Verstraete}\ \emph {et~al.}(2004)\citenamefont
  {Verstraete}, \citenamefont {Garc\'{\i}a-Ripoll},\ and\ \citenamefont
  {Cirac}}]{VerstraeteCirac2004}%
  \BibitemOpen
  \bibfield  {author} {\bibinfo {author} {\bibfnamefont {F.}~\bibnamefont
  {Verstraete}}, \bibinfo {author} {\bibfnamefont {J.~J.}\ \bibnamefont
  {Garc\'{\i}a-Ripoll}},\ and\ \bibinfo {author} {\bibfnamefont {J.~I.}\
  \bibnamefont {Cirac}},\ }\bibfield  {title} {\bibinfo {title} {Matrix product
  density operators: Simulation of finite-temperature and dissipative
  systems},\ }\href {https://doi.org/10.1103/PhysRevLett.93.207204} {\bibfield
  {journal} {\bibinfo  {journal} {Phys. Rev. Lett.}\ }\textbf {\bibinfo
  {volume} {93}},\ \bibinfo {pages} {207204} (\bibinfo {year}
  {2004})}\BibitemShut {NoStop}%
\bibitem [{\citenamefont {Parcollet}\ \emph {et~al.}(2015)\citenamefont
  {Parcollet}, \citenamefont {Ferrero}, \citenamefont {Ayral}, \citenamefont
  {Hafermann}, \citenamefont {Krivenko}, \citenamefont {Messio},\ and\
  \citenamefont {Seth}}]{ParcolletSeth2015}%
  \BibitemOpen
  \bibfield  {author} {\bibinfo {author} {\bibfnamefont {O.}~\bibnamefont
  {Parcollet}}, \bibinfo {author} {\bibfnamefont {M.}~\bibnamefont {Ferrero}},
  \bibinfo {author} {\bibfnamefont {T.}~\bibnamefont {Ayral}}, \bibinfo
  {author} {\bibfnamefont {H.}~\bibnamefont {Hafermann}}, \bibinfo {author}
  {\bibfnamefont {I.}~\bibnamefont {Krivenko}}, \bibinfo {author}
  {\bibfnamefont {L.}~\bibnamefont {Messio}},\ and\ \bibinfo {author}
  {\bibfnamefont {P.}~\bibnamefont {Seth}},\ }\bibfield  {title} {\bibinfo
  {title} {Triqs: A toolbox for research on interacting quantum systems},\
  }\href {https://doi.org/10.1016/j.cpc.2015.04.023} {\bibfield  {journal}
  {\bibinfo  {journal} {Comput. Phys. Commun.}\ }\textbf {\bibinfo {volume}
  {196}},\ \bibinfo {pages} {398} (\bibinfo {year} {2015})}\BibitemShut
  {NoStop}%
\bibitem [{\citenamefont {Seth}\ \emph {et~al.}(2016)\citenamefont {Seth},
  \citenamefont {Krivenko}, \citenamefont {Ferrero},\ and\ \citenamefont
  {Parcollet}}]{SethParcollet2016}%
  \BibitemOpen
  \bibfield  {author} {\bibinfo {author} {\bibfnamefont {P.}~\bibnamefont
  {Seth}}, \bibinfo {author} {\bibfnamefont {I.}~\bibnamefont {Krivenko}},
  \bibinfo {author} {\bibfnamefont {M.}~\bibnamefont {Ferrero}},\ and\ \bibinfo
  {author} {\bibfnamefont {O.}~\bibnamefont {Parcollet}},\ }\bibfield  {title}
  {\bibinfo {title} {Triqs/cthyb: A continuous-time quantum monte carlo
  hybridisation expansion solver for quantum impurity problems},\ }\href
  {https://doi.org/10.1016/j.cpc.2015.10.023} {\bibfield  {journal} {\bibinfo
  {journal} {Comput. Phys. Commun.}\ }\textbf {\bibinfo {volume} {200}},\
  \bibinfo {pages} {274} (\bibinfo {year} {2016})}\BibitemShut {NoStop}%
\bibitem [{\citenamefont {Mahan}(2000)}]{mahan2000-many}%
  \BibitemOpen
  \bibfield  {author} {\bibinfo {author} {\bibfnamefont {G.~D.}\ \bibnamefont
  {Mahan}},\ }\href@noop {} {\emph {\bibinfo {title} {Many-Particle Physics}}}\
  (\bibinfo  {publisher} {Springer; 3nd edition},\ \bibinfo {year}
  {2000})\BibitemShut {NoStop}%
\bibitem [{\citenamefont {Sherman}(2020)}]{Sherman2020}%
  \BibitemOpen
  \bibfield  {author} {\bibinfo {author} {\bibfnamefont {A.}~\bibnamefont
  {Sherman}},\ }\bibfield  {title} {\bibinfo {title} {Hubbard-kanamori model:
  spectral functions, negative electron compressibility, and
  susceptibilities},\ }\href {https://doi.org/10.1088/1402-4896/aba923}
  {\bibfield  {journal} {\bibinfo  {journal} {Phys. Scr.}\ }\textbf {\bibinfo
  {volume} {95}},\ \bibinfo {pages} {095804} (\bibinfo {year}
  {2020})}\BibitemShut {NoStop}%
\bibitem [{\citenamefont {Werner}\ \emph {et~al.}(2009)\citenamefont {Werner},
  \citenamefont {Gull},\ and\ \citenamefont {Millis}}]{WernerMillis2009}%
  \BibitemOpen
  \bibfield  {author} {\bibinfo {author} {\bibfnamefont {P.}~\bibnamefont
  {Werner}}, \bibinfo {author} {\bibfnamefont {E.}~\bibnamefont {Gull}},\ and\
  \bibinfo {author} {\bibfnamefont {A.~J.}\ \bibnamefont {Millis}},\ }\bibfield
   {title} {\bibinfo {title} {Metal-insulator phase diagram and orbital
  selectivity in three-orbital models with rotationally invariant hund
  coupling},\ }\href {https://doi.org/10.1103/PhysRevB.79.115119} {\bibfield
  {journal} {\bibinfo  {journal} {Phys. Rev. B}\ }\textbf {\bibinfo {volume}
  {79}},\ \bibinfo {pages} {115119} (\bibinfo {year} {2009})}\BibitemShut
  {NoStop}%
\bibitem [{\citenamefont {Makarov}\ and\ \citenamefont
  {Makri}(1994)}]{makarov1994-path}%
  \BibitemOpen
  \bibfield  {author} {\bibinfo {author} {\bibfnamefont {D.~E.}\ \bibnamefont
  {Makarov}}\ and\ \bibinfo {author} {\bibfnamefont {N.}~\bibnamefont
  {Makri}},\ }\bibfield  {title} {\bibinfo {title} {Path integrals for
  dissipative systems by tensor multiplication. condensed phase quantum
  dynamics for arbitrarily long time},\ }\href
  {https://doi.org/10.1016/0009-2614(94)00275-4} {\bibfield  {journal}
  {\bibinfo  {journal} {Chem. Phys. Lett.}\ }\textbf {\bibinfo {volume}
  {221}},\ \bibinfo {pages} {482} (\bibinfo {year} {1994})}\BibitemShut
  {NoStop}%
\bibitem [{\citenamefont {Makri}(1995)}]{makri1995-numerical}%
  \BibitemOpen
  \bibfield  {author} {\bibinfo {author} {\bibfnamefont {N.}~\bibnamefont
  {Makri}},\ }\bibfield  {title} {\bibinfo {title} {Numerical path integral
  techniques for long time dynamics of quantum dissipative systems},\ }\href
  {https://doi.org/10.1063/1.531046} {\bibfield  {journal} {\bibinfo  {journal}
  {J. Math. Phys.}\ }\textbf {\bibinfo {volume} {36}},\ \bibinfo {pages} {2430}
  (\bibinfo {year} {1995})}\BibitemShut {NoStop}%
\bibitem [{\citenamefont {Park}\ \emph {et~al.}(2024)\citenamefont {Park},
  \citenamefont {Ng}, \citenamefont {Reichman},\ and\ \citenamefont
  {Chan}}]{ParkChan2024}%
  \BibitemOpen
  \bibfield  {author} {\bibinfo {author} {\bibfnamefont {G.}~\bibnamefont
  {Park}}, \bibinfo {author} {\bibfnamefont {N.}~\bibnamefont {Ng}}, \bibinfo
  {author} {\bibfnamefont {D.~R.}\ \bibnamefont {Reichman}},\ and\ \bibinfo
  {author} {\bibfnamefont {G.~K.-L.}\ \bibnamefont {Chan}},\ }\bibfield
  {title} {\bibinfo {title} {Tensor network influence functionals in the
  continuous-time limit: Connections to quantum embedding, bath discretization,
  and higher-order time propagation},\ }\href
  {https://doi.org/10.1103/PhysRevB.110.045104} {\bibfield  {journal} {\bibinfo
   {journal} {Phys. Rev. B}\ }\textbf {\bibinfo {volume} {110}},\ \bibinfo
  {pages} {045104} (\bibinfo {year} {2024})}\BibitemShut {NoStop}%
\bibitem [{\citenamefont {Xiang}\ \emph {et~al.}(2024)\citenamefont {Xiang},
  \citenamefont {Jia}, \citenamefont {Fang},\ and\ \citenamefont
  {Li}}]{XiangLi2024}%
  \BibitemOpen
  \bibfield  {author} {\bibinfo {author} {\bibfnamefont {C.}~\bibnamefont
  {Xiang}}, \bibinfo {author} {\bibfnamefont {W.}~\bibnamefont {Jia}}, \bibinfo
  {author} {\bibfnamefont {W.-H.}\ \bibnamefont {Fang}},\ and\ \bibinfo
  {author} {\bibfnamefont {Z.}~\bibnamefont {Li}},\ }\bibfield  {title}
  {\bibinfo {title} {Distributed multi-gpu ab initio density matrix
  renormalization group algorithm with applications to the p-cluster of
  nitrogenase},\ }\href {https://doi.org/10.1021/acs.jctc.3c01228} {\bibfield
  {journal} {\bibinfo  {journal} {Journal of Chemical Theory and Computation}\
  }\textbf {\bibinfo {volume} {20}},\ \bibinfo {pages} {775} (\bibinfo {year}
  {2024})},\ \bibinfo {note} {pMID: 38198503},\ \Eprint
  {https://arxiv.org/abs/https://doi.org/10.1021/acs.jctc.3c01228}
  {https://doi.org/10.1021/acs.jctc.3c01228} \BibitemShut {NoStop}%
\bibitem [{\citenamefont {Georges}\ \emph
  {et~al.}(1996{\natexlab{b}})\citenamefont {Georges}, \citenamefont {Kotliar},
  \citenamefont {Krauth},\ and\ \citenamefont
  {Rozenberg}}]{georges1996-dynamical}%
  \BibitemOpen
  \bibfield  {author} {\bibinfo {author} {\bibfnamefont {A.}~\bibnamefont
  {Georges}}, \bibinfo {author} {\bibfnamefont {G.}~\bibnamefont {Kotliar}},
  \bibinfo {author} {\bibfnamefont {W.}~\bibnamefont {Krauth}},\ and\ \bibinfo
  {author} {\bibfnamefont {M.~J.}\ \bibnamefont {Rozenberg}},\ }\bibfield
  {title} {\bibinfo {title} {Dynamical mean-field theory of strongly correlated
  fermion systems and the limit of infinite dimensions},\ }\href
  {https://doi.org/10.1103/revmodphys.68.13} {\bibfield  {journal} {\bibinfo
  {journal} {Rev. Mod. Phys.}\ }\textbf {\bibinfo {volume} {68}},\ \bibinfo
  {pages} {13} (\bibinfo {year} {1996}{\natexlab{b}})}\BibitemShut {NoStop}%
\bibitem [{\citenamefont {Aoki}\ \emph {et~al.}(2014)\citenamefont {Aoki},
  \citenamefont {Tsuji}, \citenamefont {Eckstein}, \citenamefont {Kollar},
  \citenamefont {Oka},\ and\ \citenamefont {Werner}}]{AokiWerner2014}%
  \BibitemOpen
  \bibfield  {author} {\bibinfo {author} {\bibfnamefont {H.}~\bibnamefont
  {Aoki}}, \bibinfo {author} {\bibfnamefont {N.}~\bibnamefont {Tsuji}},
  \bibinfo {author} {\bibfnamefont {M.}~\bibnamefont {Eckstein}}, \bibinfo
  {author} {\bibfnamefont {M.}~\bibnamefont {Kollar}}, \bibinfo {author}
  {\bibfnamefont {T.}~\bibnamefont {Oka}},\ and\ \bibinfo {author}
  {\bibfnamefont {P.}~\bibnamefont {Werner}},\ }\bibfield  {title} {\bibinfo
  {title} {Nonequilibrium dynamical mean-field theory and its applications},\
  }\href {https://doi.org/10.1103/RevModPhys.86.779} {\bibfield  {journal}
  {\bibinfo  {journal} {Rev. Mod. Phys.}\ }\textbf {\bibinfo {volume} {86}},\
  \bibinfo {pages} {779} (\bibinfo {year} {2014})}\BibitemShut {NoStop}%
\bibitem [{Dat()}]{DataRepo}%
  \BibitemOpen
  \href@noop {} {\bibinfo  {journal}
  {https://github.com/guochu/GTEMPOData/tree/master/Scalable\_tensor\_network\_algorithm\_for\_quantum\_impurity\_problems}\
  }\BibitemShut {NoStop}%
\end{thebibliography}

%

\end{document}